\documentclass[aps,pre,preprint,groupedaddress,10pt,showkeys,nofootinbib]{revtex4-2}

\usepackage[driverfallback=dvipdfm]{hyperref}
\usepackage{amsmath,amssymb}
\usepackage{mathtools}
\usepackage{fancyvrb}

\allowdisplaybreaks[1]
\newcommand{\dr}{\mathrm{d}}
\usepackage{color}

\newcommand{\intn}{\mathrm{int}}
\renewcommand{\o}{\mathrm{o}}

\renewcommand{\c}{\mathrm{c}}

\newcommand{\Tr}{\mathrm{Tr}}
\newcommand{\tr}{\mathrm{tr}}

\newcommand{\rS}{\mathrm{S}}
\newcommand{\rI}{\mathrm{I}}
\newcommand{\rR}{\mathrm{R}}
\newcommand{\cl}{\mathrm{cl}}

\begin{document}


\title{Criticality in stochastic SIR model for infectious diseases}


\author{Shigehiro~Yasui}
\email[]{shigehiro.yasui@kochi-u.ac.jp}
\author{Yutaka~Hatakeyama}
\email[]{hatake@kochi-u.ac.jp}
\author{Yoshiyasu~Okuhara}
\email[]{okuharay@kochi-u.ac.jp}
\affiliation{Center of Medical Information Science, Kochi Medical School, Kochi University, Nankoku, Kochi, Japan}


\date{\today}

\begin{abstract}
We discuss the criticality in the stochastic SIR model for infectious diseases. We adopt the path-integral formalism for the propagation of infections among susceptible, infectious, and removed individuals, and perform the perturbative and nonperturbative analyses to evaluate the critical value of the basic reproduction number ${\cal R}$. In the perturbation theory, we calculate the mean values and the variances of the number of infectious individuals near the initial time, and find that the critical value ${\cal R}_{\c}=1/3$--$2/3$ should be adopted in order to suppress the stochastic spread of infections sufficiently. In the nonperturbative approach, we derive the effective potential by integrating out the stochastic fluctuations, and obtain the effective Euler-Lagrange equations for the time-evolution of the numbers of susceptible, infectious, and removed individuals.
From the asymptotic behaviors for a long time, we find that the critical value ${\cal R}_{\c}=2/3$ should be adopted for the sufficient convergence of infections.
We also find that the endemic state can be generated dynamically by the stochastic fluctuation which is absent in the conventional SIR model.
Those analyses show that the critical value of the basic reproduction number should be less than one, against the usually known critical value ${\cal R}_{\c}=1$, when the stochastic fluctuations are taken into account in the SIR model.
\end{abstract}


\maketitle

\newpage



\section{Introduction} \label{sec:introduction}

In the midst of the pandemic of Covid-19 infection since 2019~\cite{Zhu2020},
much attention is attracted to mathematical models of infectious diseases in theoretical epidemiology (see e.g. Refs.~\cite{bailey1975mathematical,anderson1992infectious,diekmann2000mathematical,daley2001epidemic}).
Currently there is a growing momentum for a multidimensional understanding that goes beyond traditional theoretical epidemiology. Many mathematical models used in theoretical epidemiology of infectious diseases are deterministic compartment models such as SIR, SIS, and SEIR
models, which stem from the Kermack-McKendrick models.\footnote{Their papers were published in 1927, 1932, and 1933, and have been later republished due to their seminal importance in 1991~\cite{Kermack1991_1,Kermack1991_2,Kermack1991_3}.} In these models, the differential equations uniquely determine the number of susceptible, infected, and removed individuals at a given time, and the number of each at a later time. On the other hand, since the transmission of infectious diseases is a stochastic phenomenon, formulations based on stochastic models have been studied since early days by one of the seminal paper by McKendrick~\cite{McKendrick1925} and have been developed as the chain bimodal model by Reed and Frost~(see e.g. Ref.~\cite{daley2001epidemic}).
One of the major advantages of considering a stochastic model over a deterministic model is that it allows us to evaluate the probabilistic confidence level of the predicted number of infected persons.
The assessment of the probabilistic uncertainty of the predicted number of infected people in the future provides an important indicator for determining epidemic control measures such as behavior change and vaccination. Especially in the early stages of infection when the number of infected people is small, or in areas with a small population, the variance due to stochastic fluctuations  cannot be ignored against the mean value.

The stochastic models that have been considered in the literature can be broadly classified into (i) discrete-time branching processes (e.g., Bienamy\'{e}-Galton-Watson branching process), (ii) continuous-time branching processes (e.g., master equation and Chapman-Kolmogorov equation), (iii) stochastic differential equations (e.g., Langevin equation and Fokker-Planck equation). They are often discussed as Markov processes, although these stochastic processes can be extended to non-Markov processes in principle. Discrete-time branching processes have been studied in infectious disease epidemiology since early days~\cite{Otter1949}, and have been utilized in actual epidemiological evaluations~\cite{Lloyd-Smith2005,Jacob2010} (see also e.g. Refs.~\cite{andersson2012stochastic,diekmann2000mathematical,daley2001epidemic,haccou2005branching}).
It was shown that there is a connection between the Reed-Frost epidemic model and the SIR model in large population limit~\cite{vonbahr_martin-lof_1980}, and the asymptotic scaling law of the number of infected persons was obtained~\cite{Martin-Lof1998,Aldous1997} (see also Ref.~\cite{dawson2017introductory}).
On the other hand, the master equation as a continuous time branching process has been mainly used in theoretical studies~\cite{Bartlett1956,daley2001epidemic,Trapman2004}. The exact solutions of the stochastic SIR model were reported for the epidemics on the networks composed of linear-chain graphs~\cite{schutz2008exact} and tree graphs~\cite{Sharkey2015}. In general cases, however, finding exact solutions of the master equation is not easy due to the nonlinearity in the SIR model without relying on the numerical calculations~\cite{Grenfell1992,Keeling2008,Jenkinson2012}.
The stochastic differential equation has mathematical research examples~\cite{dolgoarshinnykh_lalley_2006,Lalley2009,Lalley2010,Lalley2014}, whose application toward the actual epidemiology is developing~\cite{Maki2013}.

In this study, as a basic research for theoretical epidemiology of infectious diseases, we analyze the stochastic SIR model to evaluate the uncertainty of predicted number of susceptible, infected, and removed individuals due to stochastic fluctuations beyond the deterministic SIR model.
 We will discuss the impact of stochastic fluctuations on the deterministic trajectory of the deterministic SIR model (i.e., the time evolution of the number of susceptible, infected, and removed individuals) and evaluate the uncertainty by stochastic fluctuations on the predictions made by the deterministic trajectory. In the present study, the master equation of the stochastic SIR model is expressed in terms of path integrals (functional integrals) according to the Doi-Peliti formalism~\cite{Doi_1976,Peliti_1985}, utilizing field-theoretical methods stemming from quantum physics~\cite{Grassberger1982,Droz_1994,Lee1995,Cardy1996,cardy1996renormalisation,Bettelheim2001,Pastor-Satorras2001,Elgart2004,Elgart2006,Andreanov2006,Mobilia2007,Butler2009,Tauber_2011,Tauber_2012,Oizumi2013,Shih2014} (see e.g. also Refs.~\cite{Dickman2003,Janssen2005,Tauber_2005,cardy_cardy_falkovich_gawedzki_2008,Tauber2009,Tauber_2014,Weber_2017} for more details).
Interestingly, such approach was used not only for the epidemiology but also for the analysis of the cell dynamics~\cite{Zhang2014}.
The conventional deterministic SIR model corresponds to the classical trajectory in field theory, by which the deterministic SIR model can be regarded as the {\it classical} or {\it tree-level} version of the stochastic SIR model in terms of the  quantum field-theory. In this framework, stochastic fluctuations correspond to quantum fluctuations. With this setup, we focus on the basic reproduction number in the stochastic SIR model, and discuss the effect of stochastic fluctuations on the critical value of basic reproduction number ${\cal R}_{\c}$ determining the condition of convergence of infections. We show that the critical value of the basic reproduction number ${\cal R}_{c}=1$, which is known in the conventional deterministic SIR model, needs to be modified to the smaller values ${\cal R}_{c}=1/3$--$2/3$ by the influence of stochastic fluctuations. This result suggests that a more stringent quarantine regime is required for the complete control of infectious diseases.

The paper is constructed as the followings.
In Sec.~\ref{sec:formalism}, we give a formalism of the stochastic SIR model in terms of the Hamiltonian formalism and translate it to the path-integral formalism.
In Sec.~\ref{sec:results}, we present the critical value of the basic reproduction number under the influence of the stochastic fluctuations in perturbative and nonperturbative approaches.
Final section is devoted to the conclusion and the perspectives.

\section{Formalism} \label{sec:formalism}

\subsection{Hamiltonian formalism} \label{sec:Hamiltonian}
We consider
the stochastic SIR model for the infectious diseases
 in terms of the master equation~\cite{Bartlett1956,daley2001epidemic,Trapman2004}.
To begin the discussion, we define the function $P_{t}(N_{\rS},N_{\rI},N_{\rR})$ as the probability that the numbers of susceptible, infectious, and removed (recovered or dead) individuals at time $t$, which are denoted by $N_{\rS}$, $N_{\rI}$, and $N_{\rR}$, respectively.
As a stochastic process in the time-evolution of infection, we introduce the transition rates $\beta$ and $\gamma$ for transmission of the infectious diseases among people in a unit time.
Here $\beta$ indicates the rate that one susceptible person meets one infectious person and becomes infectious,
and $\gamma$ the rate that one infectious person gets removed.
In the SIR model, it is assumed that the removed person acquires immunity and never becomes susceptible again.
Considering the simple situation that the transmission of infections and the recovery from infections (or death) can occur constantly at any time, we describe the infectious dynamics  by the stochastic process through the following master equation,
\begin{align}
   \frac{{\dr}}{{\dr} t} P_{t}(N_{\rS},N_{\rI},N_{\rR})
&=
    \beta (N_{\rS}+1)(N_{\rI}-1) P_{t}(N_{\rS}+1,N_{\rI}-1,N_{\rR})
+ \gamma(N_{\rI}+1) P_{t}(N_{\rS},N_{\rI}+1,N_{\rR}-1)
   \nonumber \\ & 
 - (\beta N_{\rS}N_{\rI}+\gamma N_{\rI}) P_{t}(N_{\rS},N_{\rI},N_{\rR}),
\label{eq:SIR_master}
\end{align}
as a differential equation with a first-order derivative for time.\footnote{This is called the Kolmogorov forward equation in Ref.~\cite{daley2001epidemic}, see Eq.~(3.3.3) in this reference.}
The equation~\eqref{eq:SIR_master} is a multichannel-coupled equation of many degrees of freedom with $N_{\rS}$, $N_{\rI}$, and $N_{\rR} \in \mathbb{N}$ ($\mathbb{N}$ a set of non-negative integer numbers), thus it is difficult to solve the equation in analytic form for arbitrary populations.
The numerical calculation would also be cumbersome without simplifications because the maximum number of populations could reach an order of millions and more~\cite{Grenfell1992,Keeling2008,Jenkinson2012}.
In the present study, we transform the master equation~\eqref{eq:SIR_master} to the path-integral formalism by following the prescriptions in Refs.~\cite{Doi_1976,Peliti_1985} (see e.g. Ref.~\cite{RevModPhys.70.979} for a review), and 
research the properties of the stochastic SIR model by perturbative and nonperturbative methods based on the field-theoretical approaches.

To start with, we introduce the vector state at time $t$ in the infinitely-dimensional vector space, which is expressed by
\begin{align}
   |\Phi(t)\rangle
\equiv
   \sum_{N_{\rS},N_{\rI},N_{\rR}}
   P_{t}(N_{\rS},N_{\rI},N_{\rR})
   \prod_{i=\rS,\rI,\rR} (\hat{a}_{i}^{\dag})^{N_{i}} |0\rangle,
 \label{eq:state_vector_phi}
\end{align}
as a superposition of $P_{t}(N_{\rS},N_{\rI},N_{\rR})$ summing over various values of $N_{\rS}$, $N_{\rI}$, and $N_{\rR} \in \mathbb{N}$.
Notice that the summation is taken over for all the possible combinations of $N_{\rS}$, $N_{\rI}$, and $N_{\rR}$.
In this vector space, the basis is spanned by $\prod_{i=\rS,\rI,\rR} (\hat{a}_{i}^{\dag})^{N_{i}} |0\rangle$,
where $\hat{a}_{i}$ and $\hat{a}_{j}^{\dag}$ ($i$, $j=\rS$, $\rI$, and $\rR$) are bosonic annihilation and creation operators satisfying the commutation relations, $[\hat{a}_{i},\hat{a}_{j}^{\dag}]=\delta_{ij}$ and $[\hat{a}_{i},\hat{a}_{j}]=[\hat{a}_{i}^{\dag},\hat{a}_{j}^{\dag}]=0$.
We suppose that the operators act as $\hat{a}_{i}|m_{i}\rangle=m_{i}|m_{i}-1\rangle$ and $\hat{a}_{i}^{\dag}|m_{i}\rangle=|m_{i}+1\rangle$ for $|m_{i}\rangle$.
Here $|m_{i}\rangle$ is the state in which there exist $m_{i}$ ($\in \mathbb{N}$) individuals for $i=\rS$, $\rI$, and $\rR$.
Particularly $|0\rangle$ is defined to be the empty state in which there exists no individual.
We can regard that those annihilation and creation operators resemble essentially the ones used in quantum physics, although the normalization constants of the states are different from the conventional ones~\cite{Doi_1976,Peliti_1985}.
With the above setting, we find that the master equation~\eqref{eq:SIR_master} is transformed to the Schr\"odinger-type equation
\begin{align}
  - \frac{{\dr}}{{\dr} t} |\Phi(t)\rangle = \hat{H} |\Phi(t)\rangle,
\label{eq:Schroedinger_equation}
\end{align}
with the Hamiltonian operator defined by
\begin{align}
   \hat{H}
\equiv
   \beta
   \bigl(
      -\hat{a}_{\rS}(\hat{a}_{\rI}^{\dag})^{2}\hat{a}_{\rI}
      +\hat{a}_{\rS}^{\dag} \hat{a}_{\rS} \hat{a}_{\rI}^{\dag} \hat{a}_{\rI}
   \bigr)
+ \gamma
   \bigl(
      -\hat{a}_{\rI} \hat{a}_{\rR}^{\dag}+\hat{a}_{\rI}^{\dag}\hat{a}_{\rI}
   \bigr),
\label{eq:Hamiltonian_def}
\end{align}
in the Hamiltonian formalism.
This is just another expression of Eq.~\eqref{eq:SIR_master}.
One of the advantages of this rewriting is that
 we obtain the formal solution expressed by $|\Phi(t)\rangle = e^{-\hat{H}t}|\Phi(0)\rangle$.
In this formalism, we express the mean value of an arbitrary function $A(N_{\rS},N_{\rI},N_{\rR})$ with variables $N_{i}$ ($i=\rS$, $\rI$, and $\rR$) by
\begin{align}
   \bar{A}_{t}
&\equiv
   \sum_{N_{\rS},N_{\rI},N_{\rR}}
   P_{t}(N_{\rS},N_{\rI},N_{\rR}) A(N_{\rS},N_{\rI},N_{\rR})
\nonumber \\ 
&=
   \langle{\cal P}| A(\hat{a}_{\rS},\hat{a}_{\rI},\hat{a}_{\rR})
   |\Phi(t)\rangle,
\label{eq:expectation_value_A_operator}
\end{align}
with the projected state defined by
\begin{align}
   \langle{\cal P}|
\equiv
   \langle0|\exp\Biggl(\sum_{i=\rS,\rI,\rR}\hat{a}_{i}\Biggr).
\end{align}
The projected state has the following properties:
$\langle{\cal P}| \hat{a}_{i}^{\dag}=\langle{\cal P}|$ ($i=\rS$, $\rI$, and $\rR$) as a projection operator and $\langle{\cal P}|\Phi(t)\rangle=1$ as a normalization.
In spite of the existence of the formal solution, however, it is still difficult to obtain the explicit forms of solution in the Hamiltonian formalism.

\subsection{Path-integral formalism} \label{sec:path_integral}
We further rewrite the Schr\"odinger-type equation~\eqref{eq:Schroedinger_equation} in terms of the path-integral (functional integral) formalism by following the prescriptions in Refs.~\cite{Peliti_1985,Lee1995,Pastor-Satorras2001,Mobilia2007,Bettelheim2001,Tauber_2012,Shih2014} (see also Refs.~\cite{Dickman2003,Tauber_2005,Tauber2009,Tauber_2014} for more information).
For this purpose, we introduce the coherent state defined by
\begin{align}
   |\phi_{i}(\tau)\rangle
= \exp\biggl(-\frac{1}{2}|\phi_{i}(\tau)|^{2}+\phi_{i}(\tau)\hat{a}_{i}^{\dag}\biggr) |0\rangle,
\label{eq:coherent_state_def}
\end{align}
for a creation operator $\hat{a}_{i}^{\dag}$ for $i=\rS$, $\rI$, and $\rR$, where $\phi_{i}(\tau)$ is a parameter value in the complex number at time $\tau$.
For brevity, we sometimes use the notation as a vector $\phi(\tau)=\bigl\{\phi_{\rS}(\tau),\phi_{\rI}(\tau),\phi_{\rR}(\tau)\bigr\}$ in the followings.
Starting from Eq.~\eqref{eq:Schroedinger_equation} and slicing the time interval $\tau \in [0,t]$, instead of Eq.~\eqref{eq:expectation_value_A_operator}, we obtain the mean value at time $t$,
\begin{align}
   \langle A(\phi(t)) \rangle
=
   {\cal N}
   \int
   {\cal D}\bar{\phi}
   {\cal D}\phi \,
   A(\phi(t))
   e^{-S[\bar{\phi},\phi]},
\label{eq:expectation_value_A}
\end{align}
for an arbitrary function $A(\phi(t))$ with a variable $\phi(t)$.
Here we have introduced the path-integral on the functional of $\phi(\tau)$.
We have defined the action for $\phi$ and $\bar{\phi}$,
\begin{align}
   S[\bar{\phi},\phi] \equiv S_{0}[\bar{\phi},\phi] + S_{\intn}[\bar{\phi},\phi],
\label{eq:SIR_action}
\end{align}
which consists of the free part
\begin{align}
   S_{0}[\bar{\phi},\phi]
\equiv
    \int_{0}^{t} \dr \tau
    \biggl(
         \bar{S} \frac{{\dr}}{{\dr} \tau} S
      + \bar{I} \frac{{\dr}}{{\dr} \tau} I
      + \bar{R} \frac{{\dr}}{{\dr} \tau} R
    \biggr)
+ \bar{S}(0) \bigl(S(0)-\bar{n}_{S0}\bigr)
+ \bar{I}(0) \bigl(I(0)-\bar{n}_{I0}\bigr)
+ \bar{R}(0) \bigl(R(0)-\bar{n}_{R0}\bigr),
\label{eq:action_free}
\end{align}
and the interaction part
\begin{align}
   S_{\intn}[\bar{\phi},\phi]
\equiv
   \int_{0}^{t} \dr \tau
   \Bigl(
         \beta \bigl( - \bar{I}^{2} + \bar{S} \bar{I} - \bar{I} + \bar{S} \bigr) S I
      + \gamma \bigl( - \bar{R} + \bar{I} \bigr)I
   \Bigr).
\label{eq:action_interaction}
\end{align}
This interaction term describes the dynamics of the transmission of infections and the recovery from infections (or death).
We have used the notations $\phi=\bigl\{S,I,R\bigr\}$ instead of $\bigl\{\phi_{\rS},\phi_{\rI},\phi_{\rR}\bigr\}$ for clarity of the equations.
Here $\bar{\phi}=\bigl\{\bar{S},\bar{I},\bar{R}\bigr\}$ is a conjugate field of $\phi$.
We have introduced ${\cal N}$ for an overall constant, which is however irrelevant in the following discussions, and hence it can be discarded.
The derivations of Eq.~\eqref{eq:expectation_value_A} is quite analogous to that used in quantum field theory.
In Eq.~\eqref{eq:action_free}, we have introduced parameters $\bar{n}_{\rS0}$, $\bar{n}_{\rI0}$, and $\bar{n}_{\rR0}$ for the average numbers for the susceptible, infectious, and removed individuals at the initial stage of the time-evolution of infections ($\tau=0$), since the Poisson distribution with parameters $\bar{n}_{i0}$ ($i=\rS$, $\rI$, and $\rR$) is assumed at the initial time.
The values of $\bar{n}_{i0}$ need to be given directly from the data in real-world.
One might consider that the path-integral formalism~\eqref{eq:expectation_value_A} is much different from the original master equation~\eqref{eq:SIR_master} or the Hamiltonian formalism~\eqref{eq:Schroedinger_equation}, but they are equivalently the same.
The advantage for the path-integral formalism is to enable us the systematic and analytical methods both in perturbative and nonperturbative approaches.

In order to evaluate the path-integral in Eq.~\eqref{eq:expectation_value_A} systematically, we introduce the generating functional defined by
\begin{align}
   Z[\bar{j},j]
&\equiv
   {\cal N}
   \int {\cal D}\bar{\phi} {\cal D}\phi
   \exp
   \biggl(
        - S[\bar{\phi},\phi]
       + \int_{0}^{t} \dr \tau (\bar{\phi}j+\bar{j}\phi)
   \biggr),
\label{eq:SIR_generating_functional_source_term_def}
\end{align}
as a functional of the source functions $j=\bigl\{\sigma,\iota,\rho\bigr\}$ and $\bar{j}=\bigl\{\bar{\sigma},\bar{\iota},\bar{\rho}\bigr\}$.
The source functions are coupled to $\bar{\phi}$ and $\phi$ through $\bar{\phi}j=\bar{S}\sigma+\bar{I}\iota+\bar{R}\rho$ and $\bar{j}\phi=\bar{\sigma}S+\bar{\iota}I+\bar{\rho}R$.
To remove the reducible diagrams in $Z[\bar{j},j]$, furthermore, we define
$
   W[\bar{j},j]
= \ln Z[\bar{j},j]
$
which generates the irreducible diagrams only.
The practical method for calculating the generating functional is presented in Sec.~\ref{sec:generating_functional} in detail.
Finally,
we calculate the first- and second-order derivatives of $W[\bar{j},j]$ with respect to the source functions
to obtain the mean values of $S(t)$, $I(t)$, and $R(t)$
\begin{align}
   \langle \phi(t) \rangle
&=
   \frac{\delta W[\bar{j},j]}{\delta \bar{j}(t)} \biggr|_{j=\bar{j}=0},
\label{eq:number_mean_value_SIR}
\end{align}
and their variances
\begin{align}
   \langle (\delta \phi(t))^{2} \rangle
&=
   \frac{\delta^{2} W[\bar{j},j]}{\delta \bar{j}(t)^{2}} \biggr|_{j=\bar{j}=0}
+ \frac{\delta W[\bar{j},j]}{\delta \bar{j}(t)} \biggr|_{j=\bar{j}=0},
\label{eq:number_fluctuation_value_SIR}
\end{align}
defined by $\langle (\delta \phi(t))^{2} \rangle=\langle \phi(t)^{2} \rangle-\langle \phi(t) \rangle^{2}$ for each pair of $(\phi,\bar{j})=(S,\bar{\sigma})$, $(I,\bar{\iota})$, and $(R,\bar{\rho})$.
The derivation of Eqs.~\eqref{eq:number_mean_value_SIR} and \eqref{eq:number_fluctuation_value_SIR} are shown in Sec.~\ref{sec:expectation_values_variances}.
Notice that the equations~\eqref{eq:number_mean_value_SIR} and \eqref{eq:number_fluctuation_value_SIR} are rigorous and generally applicable.
The perturbative expansion with respect to $\beta$ and $\gamma$ can be performed straightforwardly as shown in Sec.~\ref{sec:perturbation_tree}.

\subsection{Effective action and Euler-Lagrange equations} \label{sec:effective_action_EL_equation}

For the generating functional~\eqref{eq:SIR_generating_functional_source_term_def}, beyond the perturbation theory, we adopt the nonperturbative approach by separating $\phi$ ($\bar{\phi}$) to classical parts and fluctuation parts.
By integrating the lowest-order terms for the fluctuation parts, we can obtain the effective action including the stochastic fluctuations at infrared (long wavelength) scale.
Here we represent the classical parts by $\phi_{\c}=\bigl\{S_{\c},I_{\c},R_{\c}\bigr\}$ and $\bar{\phi}_{\c}=\bigl\{\bar{S}_{\c},\bar{I}_{\c},\bar{R}_{\c}\bigr\}$ for the classical paths.
We also represent the fluctuation parts by $\phi_{\varphi}=\bigl\{S_{\varphi},I_{\varphi},R_{\varphi}\bigr\}$ and $\bar{\phi}_{\varphi}=\bigl\{\bar{S}_{\varphi},\bar{I}_{\varphi},\bar{R}_{\varphi}\bigr\}$ for the deviations
 from the classical paths, $\phi_{\c}$ and $\bar{\phi}_{\c}$.
We use ``classical" because $\phi_{\c}$ and $\bar{\phi}_{\c}$ obey the conventional differential equations in the deterministic SIR model, see Eqs.~\eqref{eq:SIR_model_differential_equation_boundary_S}-\eqref{eq:SIR_model_differential_equation_boundary_R} in Sec.~\ref{sec:EL_equation_loop}.
Using the decomposition
$
   \phi = \phi_{\c} + \phi_{\varphi}
$
and
$
   \bar{\phi} = \bar{\phi}_{\c} + \bar{\phi}_{\varphi}
$,
we rewrite the generating functional~\eqref{eq:SIR_generating_functional_source_term_def} by
\begin{align}
   Z[\bar{\phi}_{\c},\phi_{\c};\bar{j},j]
&=
   {\cal N}
   e^{-I_{\c}[\bar{\phi}_{\c},\phi_{\c}]+\bar{j}\phi_{\c}+\bar{\phi}_{c}j}
   \int
   {\cal D}\bar{\phi}_{\varphi}
   {\cal D}\phi_{\varphi} \,
   e^{-I[\bar{\phi}_{\c},\phi_{\c};\bar{\phi}_{\varphi},\phi_{\varphi}]},
\label{eq:generating_functional_WKB_0}
\end{align}
with the shorthand notation
\begin{align}
   \bar{j} \phi_{\c} + \bar{\phi}_{c} j
&\equiv
   \int_{0}^{t} \dr \tau
   \bigl( \bar{j}(\tau) \phi_{\c}(\tau) + \bar{\phi}_{c}(\tau) j(\tau) \bigr).
\label{eq:SIR_fluctuation_source_def}
\end{align}
Thus the action has been divided into the classical part, $I_{\c}[\bar{\phi}_{\c},\phi_{\c}]$, and the fluctuation part, $I[\bar{\phi}_{\c}.\phi_{\c};\bar{\phi}_{\varphi},\phi_{\varphi}]$.
Here $I_{\c}[\bar{\phi}_{\c},\phi_{\c}]$ is given by
\begin{align}
   I_{\c}[\bar{\phi}_{\c},\phi_{\c}]
&\equiv
   \int_{0}^{t} \dr \tau
   \biggl(
               \bar{S}_{\c} \frac{{\dr}}{{\dr} \tau} S_{\c}
            + \bar{I}_{\c} \frac{{\dr}}{{\dr} \tau} I_{\c}
            + \bar{R}_{\c} \frac{{\dr}}{{\dr} \tau} R_{\c}
            + \beta \bigl( - \bar{I}_{\c}^{2} + \bar{S}_{\c} \bar{I}_{\c} - \bar{I}_{\c} + \bar{S}_{\c} \bigr) S_{\c} I_{\c}
            + \gamma \bigl( - \bar{R}_{\c} + \bar{I}_{\c} \bigr)I_{\c}
               \nonumber \\ & \hspace{4em} 
            + \Bigl(
                     \bar{S}_{\c} \bigl(S_{\c}-\bar{n}_{S0}\bigr)
                  + \bar{I}_{\c} \bigl(I_{\c}-\bar{n}_{I0}\bigr)
                  + \bar{R}_{\c} \bigl(R_{\c}-\bar{n}_{R0}\bigr)
               \Bigr)
               \delta(\tau)
   \biggr),
\label{eq:SIR_fluctuation_Ic_def}
\end{align}
and $I[\bar{\phi}_{\c},\phi_{\c};\bar{\phi}_{\varphi},\phi_{\varphi}]$ is by
\begin{align}
   I[\bar{\phi}_{\c},\phi_{\c};\bar{\phi}_{\varphi},\phi_{\varphi}]
&\equiv
   \int_{0}^{t} \dr \tau \,
   \bar{\phi}_{\varphi}
   \biggl(
         \frac{{\dr}}{{\dr} \tau}
       - {\cal M}_{\c}(\tau)
      + \delta(\tau)
   \biggr)
   \phi_{\varphi},
\label{eq:SIR_fluctuation_I_def}
\end{align}
with the matrix defined by
\begin{align}
   {\cal M}_{\c}(\tau)
\equiv
   \left(
   \begin{array}{ccc}
    -\beta \bigl(\bar{I}_{\c}+1\bigr) I_{\c} & -\beta \bigl(\bar{I}_{\c}+1\bigr) S_{\c} & 0 \\
    \beta \bigl(2\bar{I}_{\c}-\bar{S}_{c}+1\bigr) I_{\c} & \beta \bigl(2\bar{I}_{\c}-\bar{S}_{\c}+1\bigr) S_{\c} - \gamma & 0 \\
    0 & \gamma & 0  
   \end{array}
   \right).
\label{eq:SIR_Mc_def}
\end{align}
As the lowest-order approximation in the fluctuation part, we have left only the bilinear terms of $\phi_{\varphi}$ and $\bar{\phi}_{\varphi}$ by regarding the fluctuation amplitudes small sufficiently.
Then, because the path-integral is reduced to the Gaussian integral,
 we can perform the path-integral analytically in Eq.~\eqref{eq:generating_functional_WKB_0}, see Sec.~\ref{sec:effective_potential_calc} for details.
Such approximation is essentially the same as the Wentzel-Kramers-Brillouin (WKB) approximation in quantum field theory.

As in Sec.~\ref{sec:path_integral}, furthermore, we define the generating functional $W[\bar{\phi}_{\c},\phi_{\c};\bar{j},j] = \ln Z[\bar{\phi}_{\c},\phi_{\c};\bar{j},j]$ for Eq.~\eqref{eq:generating_functional_WKB_0} and calculate the irreducible diagrams only.
Performing the path-integrals for $\phi_{\varphi}$ and $\bar{\phi}_{\varphi}$, we obtain the effective action
\begin{align}
   W[\bar{\phi}_{\c},\phi_{\c};\bar{j},j]
=
 - I_{\c}[\bar{\phi}_{\c},\phi_{\c}]
+ \bar{j}\phi_{\c}+\bar{\phi}_{c}j
 - \frac{1}{2} \Tr \ln \Delta^{-1}[\bar{\phi}_{\c},\phi_{\c}],
\label{eq:SIR_effective_potential_1loop}
\end{align}
where $\Delta^{-1}[\bar{\phi}_{\c},\phi_{\c}]$ is defined by
\begin{align}
   \Delta^{-1}[\bar{\phi}_{\c},\phi_{\c}](\tau_{1},\tau_{2})
\equiv
   \delta(\tau_{2}-\tau_{1})
   \biggl(
         \frac{{\dr}}{{\dr} \tau_{1}}
       - {\cal M}_{\c}(\tau_{1})
      + \delta(\tau_{1})
   \biggr),
\end{align}
as a propagator of the fluctuation fields between two time points $\tau_{1}$ and $\tau_{2}$.
We denote the trace by $\displaystyle \Tr \, F = \int_{0}^{t} \dr \tau \, F(\tau,\tau)$ as an integral over the time for the two-point function $F(\tau_{1},\tau_{2})$.
Notice that $W[\bar{\phi}_{\c},\phi_{\c};\bar{j},j]$ is called the effective action since the fluctuation parts are included in the logarithmic term in the last.
In Eq.~\eqref{eq:SIR_effective_potential_1loop}, finally, we apply the Legendre transformation to obtain the effective potential defined by
\begin{align}
   \Gamma[\bar{\phi},\phi]
\equiv
   W[\bar{\phi}_{\c},\phi_{\c};\bar{j},j]
 - \bar{j}\phi
 - \bar{\phi}j,
\label{eq:effective_potential_def}
\end{align}
with the variables defined by
\begin{align}
   \phi \equiv \frac{\delta W}{\delta \bar{j}}[\bar{\phi}_{\c},\phi_{\c};\bar{j},j], \quad
   \bar{\phi} \equiv \frac{\delta W}{\delta j}[\bar{\phi}_{\c},\phi_{\c};\bar{j},j].
\label{eq:effective_field_def}
\end{align}
Here $\phi=\bigl\{S,R,I\bigr\}$ and $\bar{\phi}=\bigl\{\bar{S},\bar{R},\bar{I}\bigr\}$ should be distinguished from the integral variables used in Sec.~\ref{sec:path_integral}, although the same symbols are used for saving the letters.
Evaluating $\phi$ and $\bar{\phi}$ in Eq.~\eqref{eq:effective_field_def} by using the generating functional~\eqref{eq:generating_functional_WKB_0},
we find $\phi \approx \phi_{\c}$ and $\bar{\phi} \approx \bar{\phi}_{\c}$ at the lowest approximation in the loop expansion.
Thus, we obtain the effective Euler-Lagrange (EL) equations for $\phi$,
\begin{align}
   \frac{\delta}{\delta \bar{S}} \Gamma[\bar{\phi},\phi] \biggr|_{\bar{\phi}=0}
=
   \frac{\delta}{\delta \bar{I}} \Gamma[\bar{\phi},\phi] \biggr|_{\bar{\phi}=0}
=
   \frac{\delta}{\delta \bar{R}} \Gamma[\bar{\phi},\phi] \biggr|_{\bar{\phi}=0}
=0,
\label{eq:effective_EL_equations}
\end{align}
which give the effective paths modified by the stochastic fluctuations at the lowest order around the classical paths.

\section{Results} \label{sec:results}

With the setup in Sec.~\ref{sec:formalism}, we calculate various quantities (mean-values, variances, and so on) related to the numbers of susceptible, infectious, and removed individuals.
Here we show the results on the basic reproduction number which is critical to the watershed of spread and convergence of infectious diseases. 

\subsection{Mean values and variances at classical level in perturbative expansion}
\label{sec:perturbation_tree}

We show the results of the mean values and the variances for $S(t)$, $I(t)$, and $R(t)$ given in Eqs.~\eqref{eq:number_mean_value_SIR} and \eqref{eq:number_fluctuation_value_SIR}.
Besides the mean values, the variances are important quantities to estimate how much uncertainty can exist in the predicted number of susceptible, infectious, and removed individuals.
In the calculation, we adopt the perturbation as a series of $t$ up to and including ${\cal O}(t^{2})$ at classical level.
As results, we obtain
\begin{align}
   \langle S(t) \rangle_{\cl_{\cl}}
&=
   \bar{n}_{S0}
 - \beta
    \bar{n}_{S0} \bar{n}_{I0}
   t
+ \biggl(
         \frac{1}{2} \beta^{2}
         \bigl(
             - \bar{n}_{S0}^{2} \bar{n}_{I0}
            + \bar{n}_{S0} \bar{n}_{I0}^{2}
         \bigr)
      + \frac{1}{2} \beta \gamma \bar{n}_{S0} \bar{n}_{I0}
   \biggr)
   t^{2}
+ {\cal O}(t^{3}),
\label{eq:SIR_S_average_tree}
\\ 
   \langle I(t) \rangle_{\cl}
&=
   \bar{n}_{I0}
+ \bigl( \beta \bar{n}_{S0} - \gamma \bigr)
   \bar{n}_{I0} t
+ \biggl(
         \frac{1}{2} \beta^{2}
         \bigl( \bar{n}_{S0} - \bar{n}_{I0} \bigr) \bar{n}_{S0}
       - \beta \gamma \bar{n}_{S0}
      + \frac{1}{2} \gamma^{2}
   \biggr)
   \bar{n}_{I0}
   t^{2}
+ {\cal O}(t^{3}),
\label{eq:SIR_I_average_tree}
\\ 
   \langle R(t) \rangle_{\cl}
&=
   \bar{n}_{R0}
+ \gamma \bar{n}_{I0} t
+ \biggl(
         \frac{1}{2}
         \beta \gamma
         \bar{n}_{S0} \bar{n}_{I0}
       - \frac{1}{2} \gamma^{2}
         \bar{n}_{I0}
   \biggr)
   t^{2}
+ {\cal O}(t^{3}),
\label{eq:SIR_R_average_tree}
\end{align}
for the mean values and
\begin{align}
   \bigl\langle \bigl(\delta R(t)\bigr)^{2} \bigr\rangle_{\cl}
&=
   \bar{n}_{S0}
 - \beta
   \bar{n}_{S0} \bar{n}_{I0}
   t
+ \biggl(
         \frac{1}{2} \beta^{2}
         \bigl(
               \bar{n}_{S0}^{2} \bar{n}_{I0}
            + \bar{n}_{S0} \bar{n}_{I0}^{2}
         \bigr)
      + \frac{1}{2} \beta \gamma \bar{n}_{S0} \bar{n}_{I0}
   \biggr)
   t^{2}
+ {\cal O}(t^{3}),
\label{eq:SIR_S_fluctuation_tree}
\\[0.5em] 
   \bigl\langle \bigl(\delta I(t)\bigr)^{2} \bigr\rangle_{\cl}
&=
   \bar{n}_{I0}
+ \bigl(
         3
         \beta
         \bar{n}_{S0}
       - \gamma
   \bigr)
   \bar{n}_{I0}
   t
+ \biggl(
         \frac{1}{2} \beta^{2}
         \bigl(
               7\bar{n}_{S0}^{2} \bar{n}_{I0}
             - 5\bar{n}_{S0} \bar{n}_{I0}^{2}
         \bigr)
       - 4
         \beta \gamma
         \bar{n}_{S0} \bar{n}_{I0}
      + \frac{1}{2}
         \gamma^{2}
         \bar{n}_{I0}
   \biggr)
   t^{2}
+ {\cal O}(t^{3}),
\label{eq:SIR_I_fluctuation_tree}
\\[0.5em] 
   \bigl\langle \bigl(\delta R(t)\bigr)^{2} \bigr\rangle_{\cl}
&=
   \bar{n}_{R0}
+ \gamma \bar{n}_{I0} t
+ \biggl(
         \frac{1}{2}
         \beta \gamma
         \bar{n}_{S0} \bar{n}_{I0}
       - \frac{1}{2} \gamma^{2}
         \bar{n}_{I0}
   \biggr)
   t^{2}
+ {\cal O}(t^{3}),
\label{eq:SIR_R_fluctuation_tree}
\end{align}
for the variances.
We notice that the variances at classical level  include the effect from the stochastic fluctuation, while the mean values only reproduce the results of the classical SIR model. 
In fact, it is confirmed that the sum of the mean values satisfy exactly the conservation of the total population number as
$
   \langle S(t) \rangle_{\cl} + \langle I(t) \rangle_{\cl} + \langle R(t) \rangle_{\cl}
=
   \bar{n}_{S0} + \bar{n}_{I0} + \bar{n}_{R0}
$,
and that
 $\langle S(t) \rangle_{\cl}$, $\langle I(t) \rangle_{\cl}$, and $\langle R(t) \rangle_{\cl}$
 coincide with the solutions from the classical SIR model, see Eqs.~\eqref{eq:SIR_model_differential_equation_boundary_S}-\eqref{eq:SIR_model_differential_equation_boundary_R}.
Noticing the relation
$
   \langle I(t) \rangle_{\cl}
=
   \bar{n}_{I0}
+ \bigl( {\cal R} - 1 \bigr)
   \gamma \bar{n}_{I0} t
+ {\cal O}(t^{2})
$
 for the basic reproduction number ${\cal R}=\beta\bar{n}_{S0}/\gamma$.
The value of ${\cal R}$ indicates the average number of infected individuals at secondary infection when the number of infected individuals in the primary infection are negligible at the early stage of the infection process.
We then read that  the number of infectious individuals starts to increase for the positive coefficient (${\cal R}>1$), while it does to decrease for the negative coefficient (${\cal R}<1$).
Thus, the critical value for the infection spreadings seems to be provided by ${\cal R}_{\c}=1$.
However, we should notice that the mean value $\langle I(t) \rangle_{\cl}$ is an average value of the populations which can occur possibly at most.
We may consider the possible situations that the number of infectious individuals could still increase even for ${\cal R}<1$,
which case is allowed to occur within the probabilistic uncertainty. 
Focusing on the uncertainty shown in Eq.~\eqref{eq:SIR_I_fluctuation_tree},
we carefully investigate the variances
$
   \bigl\langle \bigl(\delta I(t)\bigr)^{2} \bigr\rangle_{\cl}
=
   \bar{n}_{I0}
+ \bigl(
         3{\cal R}
       - 1
   \bigr)
   \gamma \bar{n}_{I0} t
+ {\cal O}(t^{2})
$
at the lowest order.
If we want request stringently the situation that the number of the infectious individuals decrease certainly for any possible situations in ${\cal R}<1$,
we need to request further condition that $\bigl\langle \bigl(\delta I(t)\bigr)^{2} \bigr\rangle_{\cl}$ should not increase as well.
We thus find another condition ${\cal R}<1/3$ in addition to the conventional condition, ${\cal R}<1$.
Therefore, we need the stronger condition ${\cal R}<1/3$, giving the smaller critical value ${\cal R}_{\c}=1/3$, in order for that the infection starts to converge certainly within the probabilistic uncertainty.

Beyond the tree-level approximation,
we can include the multi-loop diagrams in the perturbative calculation.
The calculation is tedious, but straightforwardly done by following Eqs.~\eqref{eq:number_mean_value_SIR} and \eqref{eq:number_fluctuation_value_SIR}.
The results are
\begin{align}
   \langle S(t) \rangle
&=
   \bar{n}_{S0}
+ \beta
   \biggl(
       - \frac{1}{2} \bar{n}_{S0}
       - \bar{n}_{S0} \bar{n}_{I0}
   \biggr)
   t
+ \Biggl(
         \frac{1}{2} \beta^{2}
         \biggl(
             - 2\bar{n}_{S0}^{2}
            + \frac{1}{4}\bar{n}_{S0}
            + 3\bar{n}_{S0} \bar{n}_{I0}
             - \bar{n}_{S0}^{2} \bar{n}_{I0}
            + \bar{n}_{S0} \bar{n}_{I0}^{2}
         \biggr)
      + \frac{1}{2} \beta \gamma \bar{n}_{S0} \bar{n}_{I0}
   \Biggr)
   t^{2}
   \nonumber \\ & 
+ {\cal O}(t^{3}),
\label{eq:SIR_S_average}
\\ 
   \langle I(t) \rangle
&=
   \bar{n}_{I0}
+ \Biggl(
         \beta
         \biggl(
               \frac{1}{2} \bar{n}_{S0}
             - \frac{1}{2} \bar{n}_{I0}
            + \bar{n}_{S0} \bar{n}_{I0}
         \biggr)
       - \gamma \bar{n}_{I0}
   \Biggr)
   t
   \nonumber \\ & \hspace{0em} 
+ \Biggl(
         \frac{1}{2} \beta^{2}
         \biggl(
             - \bar{n}_{S0}
            + \bar{n}_{S0}^{2}
             - 4\bar{n}_{S0} \bar{n}_{I0}
            + \frac{1}{4}\bar{n}_{I0}
            + \bar{n}_{S0}^{2} \bar{n}_{I0}
             - \bar{n}_{S0} \bar{n}_{I0}^{2}
         \biggr)
      + \beta \gamma
         \biggl(
             - \frac{1}{2}\bar{n}_{S0}
            + \frac{1}{2}\bar{n}_{I0}
             - \bar{n}_{S0} \bar{n}_{I0}
         \biggr)
      + \frac{1}{2}
         \gamma^{2}
         \bar{n}_{I0}
   \Biggr)
   t^{2}
   \nonumber \\ & \hspace{0em} 
+ {\cal O}(t^{3}),
\label{eq:SIR_I_average}
\\ 
   \langle R(t) \rangle
&=
   \bar{n}_{R0}
+ \gamma \bar{n}_{I0} t
+ \Biggl(
         \beta \gamma
         \biggl(
               \frac{1}{2}
               \bar{n}_{S0}
             - \frac{1}{4}
               \bar{n}_{I0}
            + \frac{1}{2}
               \bar{n}_{S0} \bar{n}_{I0}
         \biggr)
       - \frac{1}{2} \gamma^{2}
         \bar{n}_{I0}
   \Biggr)
   t^{2}
+ {\cal O}(t^{3}),
\label{eq:SIR_R_average}
\end{align}
for the mean values, and
\begin{align}
   \bigl\langle \bigl(\delta S(t)\bigr)^{2} \bigr\rangle
&=
   \bar{n}_{S0}
+ \beta
   \biggl(
       - \frac{1}{2} \bar{n}_{S0}
       - \bar{n}_{S0} \bar{n}_{I0}
   \biggr)
   t
+ \Biggl(
         \frac{1}{2} \beta^{2}
         \biggl(
             - 2\bar{n}_{S0}^{2}
            + \frac{1}{4}\bar{n}_{S0}
            + 3\bar{n}_{S0} \bar{n}_{I0}
            + \bar{n}_{S0}^{2} \bar{n}_{I0}
            + \bar{n}_{S0} \bar{n}_{I0}^{2}
         \biggr)
      + \frac{1}{2} \beta \gamma \bar{n}_{S0} \bar{n}_{I0}
   \Biggr)
   t^{2}
   \nonumber \\ & 
+ {\cal O}(t^{3}),
\label{eq:SIR_S_fluctuation}
\\[0.5em] 
   \bigl\langle \bigl(\delta I(t)\bigr)^{2} \bigr\rangle
&=
   \bar{n}_{I0}
+ \Biggl(
         \beta
         \biggl(
               \frac{1}{2} \bar{n}_{S0}
             - \frac{1}{2} \bar{n}_{I0}
            + 3\bar{n}_{S0} \bar{n}_{I0}
         \biggr)
       - \gamma \bar{n}_{I0}
   \Biggr)
   t
   \nonumber \\ & \hspace{0em} 
+ \Biggl(
         \frac{1}{2} \beta^{2}
         \biggl(
             - \bar{n}_{S0}
            + 2\bar{n}_{S0}^{2}
             - 12\bar{n}_{S0} \bar{n}_{I0}
            + \frac{1}{4}\bar{n}_{I0}
            + 7\bar{n}_{S0}^{2} \bar{n}_{I0}
             - 5\bar{n}_{S0} \bar{n}_{I0}^{2}
         \biggr)
      + \beta \gamma
         \biggl(
             - \frac{1}{2}\bar{n}_{S0}
            + \frac{1}{2}\bar{n}_{I0}
             - 4\bar{n}_{S0} \bar{n}_{I0}
         \biggr)
         \nonumber \\ & \hspace{2em} 
      + \frac{1}{2}
         \gamma^{2}
         \bar{n}_{I0}
   \Biggr)
   t^{2}
+ {\cal O}(t^{3}),
\label{eq:SIR_I_fluctuation}
\\[0.5em] 
   \bigl\langle \bigl(\delta R(t)\bigr)^{2} \bigr\rangle
&=
   \bar{n}_{R0}
+ \gamma \bar{n}_{I0} t
+ \Biggl(
         \beta \gamma
         \biggl(
               \frac{1}{2}
               \bar{n}_{S0}
             - \frac{1}{4}
               \bar{n}_{I0}
            + \frac{1}{2}
              \bar{n}_{S0} \bar{n}_{I0}
         \biggr)
       - \frac{1}{2} \gamma^{2}
         \bar{n}_{I0}
   \Biggr)
   t^{2}
+ {\cal O}(t^{3}),
\label{eq:SIR_R_fluctuation}
\end{align}
for the variances.
It is confirmed that the above equations are reduced to the results at tree level in Eqs.~\eqref{eq:SIR_S_average_tree}-\eqref{eq:SIR_R_fluctuation_tree} by dropping the higher-order terms of $\bar{n}_{S0}$, $\bar{n}_{I0}$, and $\bar{n}_{R0}$, as it should be expected.
We focus on the behavior of infectious individuals in $\langle I(t) \rangle$ and $\langle \bigl(\delta I(t)\bigr)^{2} \rangle$ in Eqs~\eqref{eq:SIR_I_average} and \eqref{eq:SIR_I_fluctuation} .
Introducing the approximation of $\bar{n}_{I0} \ll \bar{n}_{S0}$ and leaving the lowest-order terms, we obtain
\begin{align}
   \langle I(t) \rangle
&\approx
   \bar{n}_{I0}
+ \bar{n}_{I0}
   \biggl(
         \frac{R_{\o}}{2\bar{n}_{I0}} 
      + R_{\o}
       - 1
   \biggr)
   \gamma
   t
+ {\cal O}(t^{2}),
\label{eq:SIR_I_average_approx}
\end{align}
and
\begin{align}
   \bigl\langle \bigl(\delta I(t)\bigr)^{2} \bigr\rangle
&\approx
   \bar{n}_{I0}
+ \bar{n}_{I0}
   \biggl(
         \frac{3R_{\o}}{2\bar{n}_{I0}} 
      + R_{\o}
       - 1
   \biggr)
   \gamma
   t
+ {\cal O}(t^{2}).
\label{eq:SIR_I_fluctuation_approx}
\end{align}
The condition for the decrease of the mean value of infectious individuals is given by the critical value of the basic reproduction number ${\cal R}_{\c} = 2/3$, which is given by the second term in r.h.s. of Eq.~\eqref{eq:SIR_I_average_approx} under the condition that the basic reproduction number is defined for $\bar{n}_{I0}=1$.
Besides, similarly, the condition for the decrease of the variance is given by ${\cal R}_{\c}=2/5$.
We will show the numerical behaviors at multi-loops in Eqs.~\eqref{eq:SIR_S_average}-\eqref{eq:SIR_R_average} in comparison with the results from the nonperturbative approach in the next subsection.

\subsection{Effective Euler-Lagrange equations with nonperturbative effects} \label{sec:EL_equation_loop}

Beyond the perturbation theory, we show the results of the effective EL equations~\eqref{eq:effective_EL_equations} at one-loop level to examine the nonperturbative effects in the stochastic SIR model.
Following the procedure in Sec.~\ref{sec:effective_action_EL_equation},
we obtain the EL equations
\begin{align}
   \frac{{\dr}}{{\dr} t} S
+ \beta S I
+ (S-\bar{n}_{S0}) \delta(t)
&= 0,
\label{eq:SIR_model_differential_equation_boundary_S}
\\ 
   \frac{{\dr}}{{\dr} t} I
 - \beta S I
+ \gamma I
+ (I-\bar{n}_{I0}) \delta(t)
&= 0,
\label{eq:SIR_model_differential_equation_boundary_I}
\\ 
   \frac{{\dr}}{{\dr} t} R
 - \gamma I
+ (R-\bar{n}_{R0}) \delta(t)
&= 0,
\label{eq:SIR_model_differential_equation_boundary_R}
\end{align}
at classical level, as the conventional SIR model, and
\begin{align}
   \frac{{\dr}}{{\dr} t} S
+ \beta
   S I
 - \frac{1}{4}
   \bigl( - \beta S - \gamma \bigr)
+ \bigl(S-\bar{n}_{S0}\bigr)
   \delta(t)
&= 0,
\label{eq:SIR_EOM_fluctuation_limit_S}
\\ 
   \frac{{\dr}}{{\dr} t} I
 - \beta
   S I
+ \gamma I
 - \frac{1}{4}
   \bigl( 2\beta S - \beta I \bigr)
+ \bigl(I-\bar{n}_{I0}\bigr)
   \delta(t)
&= 0,
\label{eq:SIR_EOM_fluctuation_limit_I}
\\ 
   \frac{{\dr}}{{\dr} t} R
 - \gamma I
+ \bigl(R-\bar{n}_{R0}\bigr)
   \delta(t)
&= 0,
\label{eq:SIR_EOM_fluctuation_limit_R}
\end{align}
at one-loop level.
Notice the terms with $\delta$-functions in the above equations represent the boundary conditions at the initial time.
In Eqs.~\eqref{eq:SIR_EOM_fluctuation_limit_S}-\eqref{eq:SIR_EOM_fluctuation_limit_R}, we have included the nonperturbative effects of stochastic fluctuations by calculating the trace term in Eq.~\eqref{eq:SIR_effective_potential_1loop}.
Precisely, the terms generated by the stochastic fluctuations are dependent on the terminal time $t$ as shown in the effective potential~\eqref{eq:SIR_effective_potential_1} in Sec.~\ref{sec:effective_potential_calc}, because it is regarded as the typical scalable quantity for the spreads of infectious diseases.
In the present discussion, we focus on the long-time limit as a simple case, and regard that the above equations are applied to the time-interval $\tau \in [0,\infty)$ by setting $t \rightarrow \infty$.
In Eqs.~\eqref{eq:SIR_model_differential_equation_boundary_S}-\eqref{eq:SIR_EOM_fluctuation_limit_R}, finally, $\tau$ has been replaced to $t$ ($t\in[0,\infty)$) just for the appearance of notations. 
The set of equations~\eqref{eq:SIR_model_differential_equation_boundary_S}-\eqref{eq:SIR_model_differential_equation_boundary_R} reproduces the conventional SIR  model, and may be called the classical SIR model because they include only the tree level without the stochastic fluctuations.
The stochastic fluctuations stemming from the one-loop approximation appear in the additional terms in Eqs.~\eqref{eq:SIR_EOM_fluctuation_limit_S}-\eqref{eq:SIR_EOM_fluctuation_limit_R}, which will be shown to be important in the evaluation of the critical value of the basic reproduction number.

For investigating the nonperturbative effect concretely, we examine the behavior of $I(t)$ modified by the fluctuation term in Eq.~\eqref{eq:SIR_EOM_fluctuation_limit_I} in large $t$.
As a simple setting for small number of infectious individuals,
 we suppose that the number of infectious individuals is much less than that of the susceptible individuals, i.e., $I(t) \ll S(t)$,
and that the number of susceptible individuals
 has only negligible change,
  i.e., $S(t) \approx \bar{n}_{S0}$, through the infection process assuming large population sizes.
Then, we obtain the analytic forms of solutions
\begin{align}
   I(t)
=
   \bar{n}_{I0} e^{({\cal R}-1)\gamma t},
\label{eq:SIR_NP_simple_tree}
\end{align}
from the classical EL equation~\eqref{eq:SIR_model_differential_equation_boundary_I}, and 
\begin{align}
   I(t)
=
   \dfrac{1}{2\biggl({\cal R} - 1 - \dfrac{\beta}{4\gamma}\biggr)}
   \Biggl(
       - {\cal R}
      + \Biggl(
               2 \bar{n}_{I0} \biggl( {\cal R} -1 - \dfrac{\beta}{4\gamma} \biggr)
            + {\cal R}
         \Biggr)
         \exp
         \Biggl(
               \biggl( {\cal R} - 1 - \frac{\beta}{4\gamma} \biggr) \gamma t
         \Biggr)
   \Biggr),
\label{eq:SIR_NP_simple_one_loop}
\end{align}
from the effective EL equation~\eqref{eq:SIR_EOM_fluctuation_limit_I}.
We now consider the asymptotic value $I(\infty)$ at $t \rightarrow \infty$, and set ${\cal R}<1$ for the convergence of infections given at classical level.
In this case, the solution from the classical EL equation~\eqref{eq:SIR_NP_simple_tree} indeed gives the convergence, because $I(\infty)$ becomes zero at $t \rightarrow \infty$.
On the other hand, the solution from the effective EL equation~\eqref{eq:SIR_NP_simple_one_loop} gives
\begin{align}
    I(\infty)
\approx
   \frac{{\cal R}}{2\bigl(1-{\cal R}\bigr)},
\label{eq:SIR_NP_simple_one_loop_limit}
\end{align}
resulting in a nonzero value at $t \rightarrow \infty$.
Here we have assumed the approximation that $\bar{n}_{S0}$ is large enough and $\beta/\gamma$ can be neglected as in realistic situations.
Thus, we find that the stochastic effect modifies drastically the asymptotic state at the later time of infection process:
 the effective endemic state is dynamically induced by the stochastic fluctuations.
The appearance of the endemic state is a particular feature in the stochastic SIR model that is qualitatively different from the classical SIR model, because there is no endemic state in the conventional SIR model.

The nonperturbative result provides a constraint condition on the critical value of the reproduction number for the convergence of infections.
Let us see closely the numerical behavior in Eq.~\eqref{eq:SIR_NP_simple_one_loop_limit}.
When we choose the basic reproduction number ${\cal R}=0.95$ or $0.9$ for example as a number less than one,
it gives $I(\infty)=9.5$ or $4.5$ persons meaning that there remain $9.5$ or $4.5$ infectious individuals in average for a long time, respectively, in the effective endemic state.
If ${\cal R}$ is much less than one, we may expect that the number of infectious individuals can be much smaller.
We then may ask the condition for that the infectious individuals vanish for some small number of ${\cal R}$.
We consider that the convergence would be achieved practically when the number of infectious individuals is smaller than one: $I(\infty)<1$.
Then, as an answer to the question, we find that ${\cal R}<2/3$, i.e., the critical value ${\cal R}_{\c}=2/3$, is needed to accomplish the vanishment of infections, although it doe not give zero for $I(\infty)$ in a rigorous sense.
Under this condition, we can accomplish the convergence of infections completely, accompanying stochastic fluctuations, at any later time.
The new critical value ${\cal R}_{\c}=2/3$ is smaller than ${\cal R}_{\c}=1$ in the conventional SIR model, and hence more stringent quarantine measures should be needed when the stochastic fluctuations are taken into account.
This critical value seems to be consistent with ${\cal R}_{\c}=1/3$--$2/3$ obtained by the perturbative calculation in Sec.~\ref{sec:perturbation_tree}.
We remark that
 the value of $I(\infty)$ becomes divergent when ${\cal R}$ approaches one from the below.
However, 
 ${\cal R}$ should not exceed one in the present approximation, because we have assumed the
  situation that the number of infectious individuals is much less than that of susceptible individuals.

\begin{figure}
\includegraphics[keepaspectratio, scale=0.2]{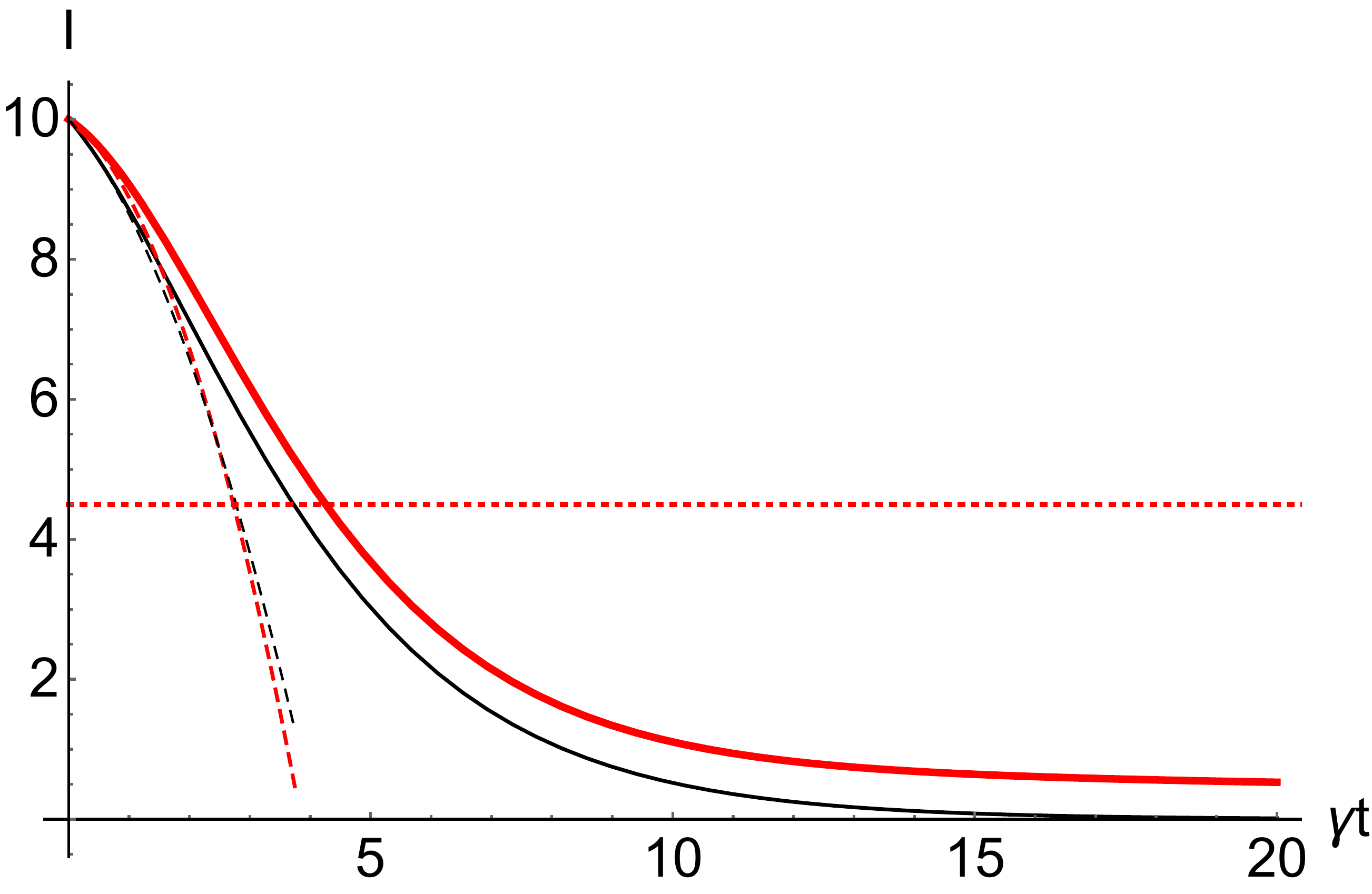}%
\includegraphics[keepaspectratio, scale=0.2]{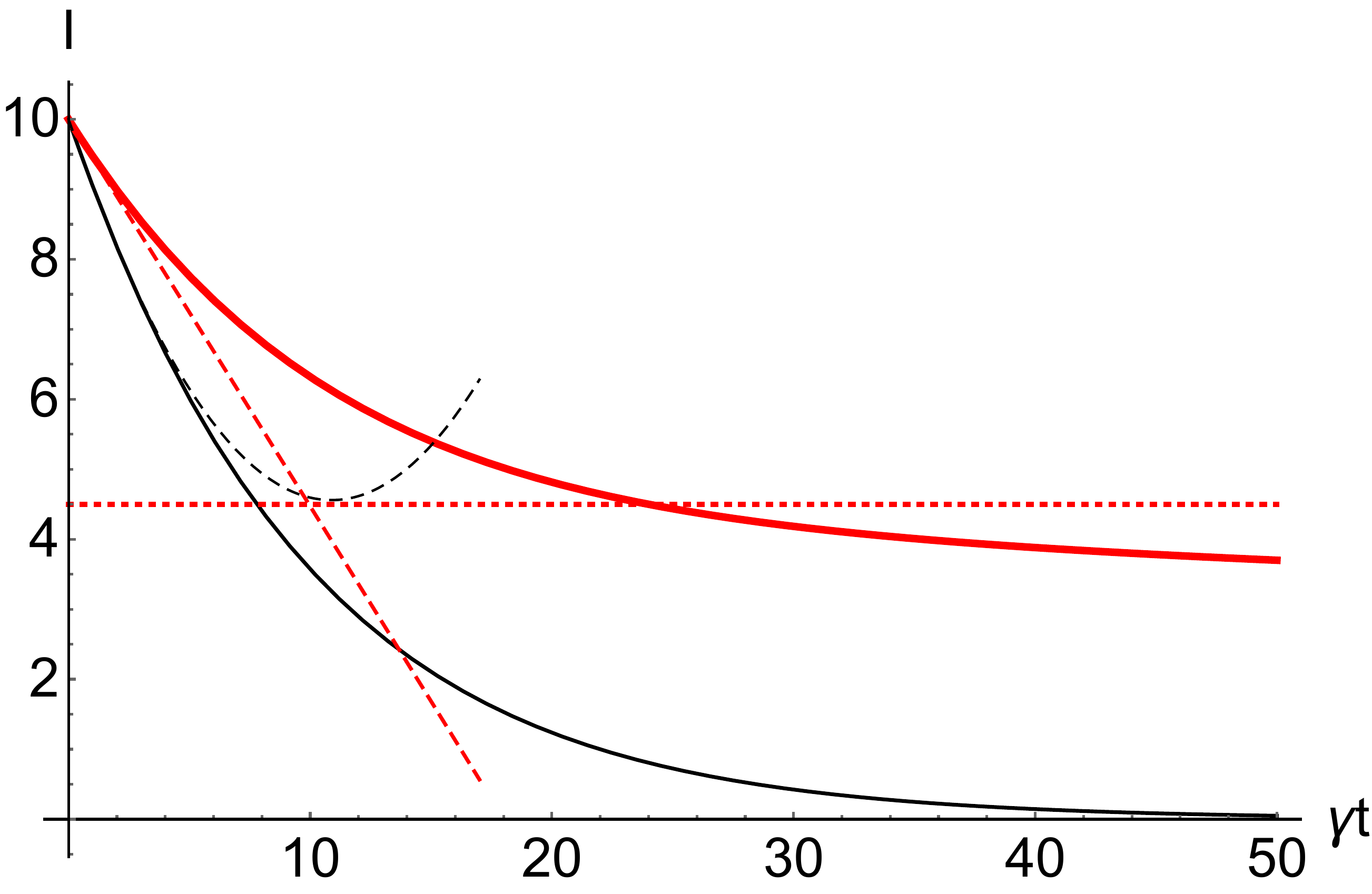}%
\includegraphics[keepaspectratio, scale=0.2]{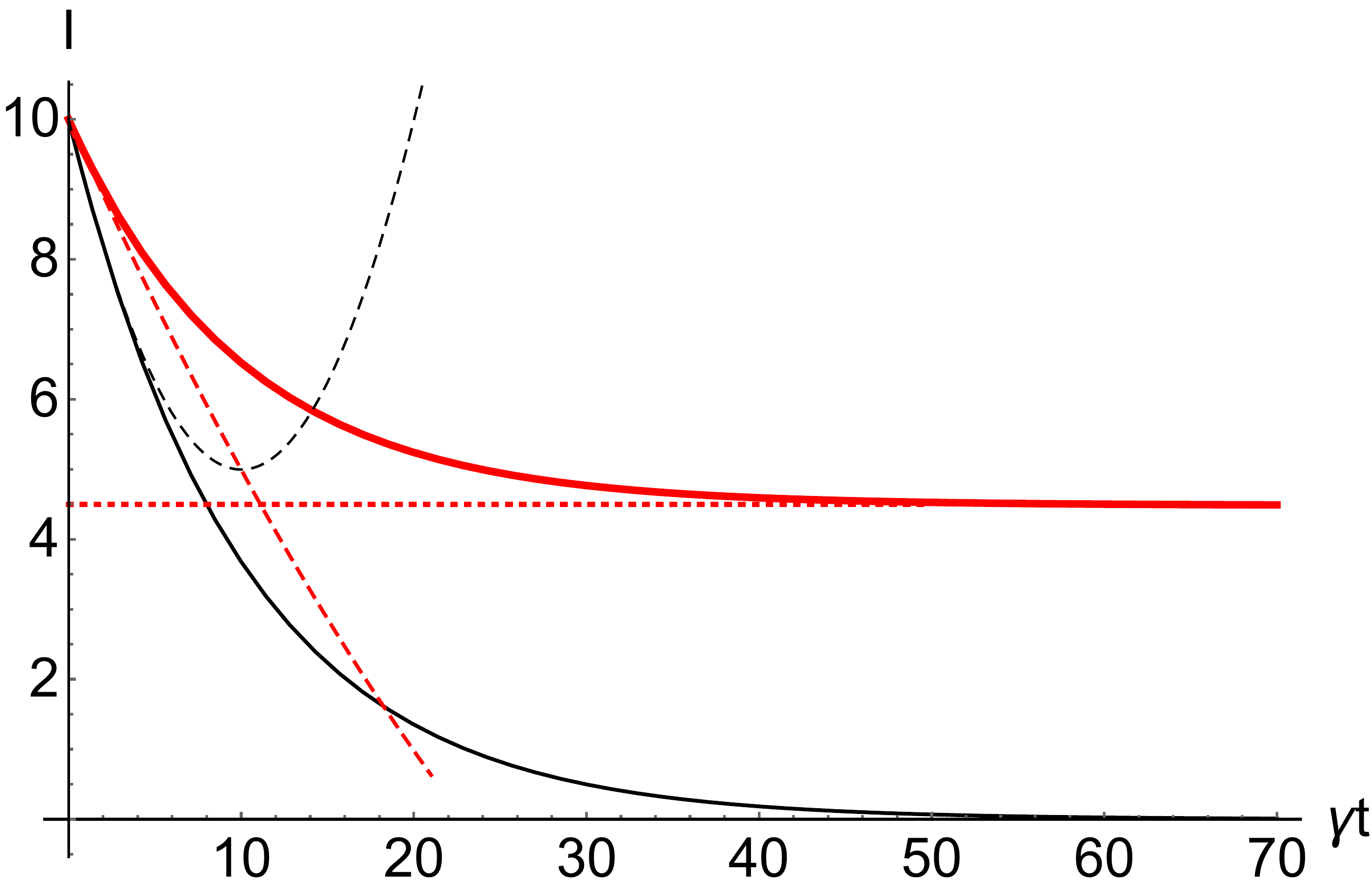}%
\\
\vspace*{2em}
\includegraphics[keepaspectratio, scale=0.2]{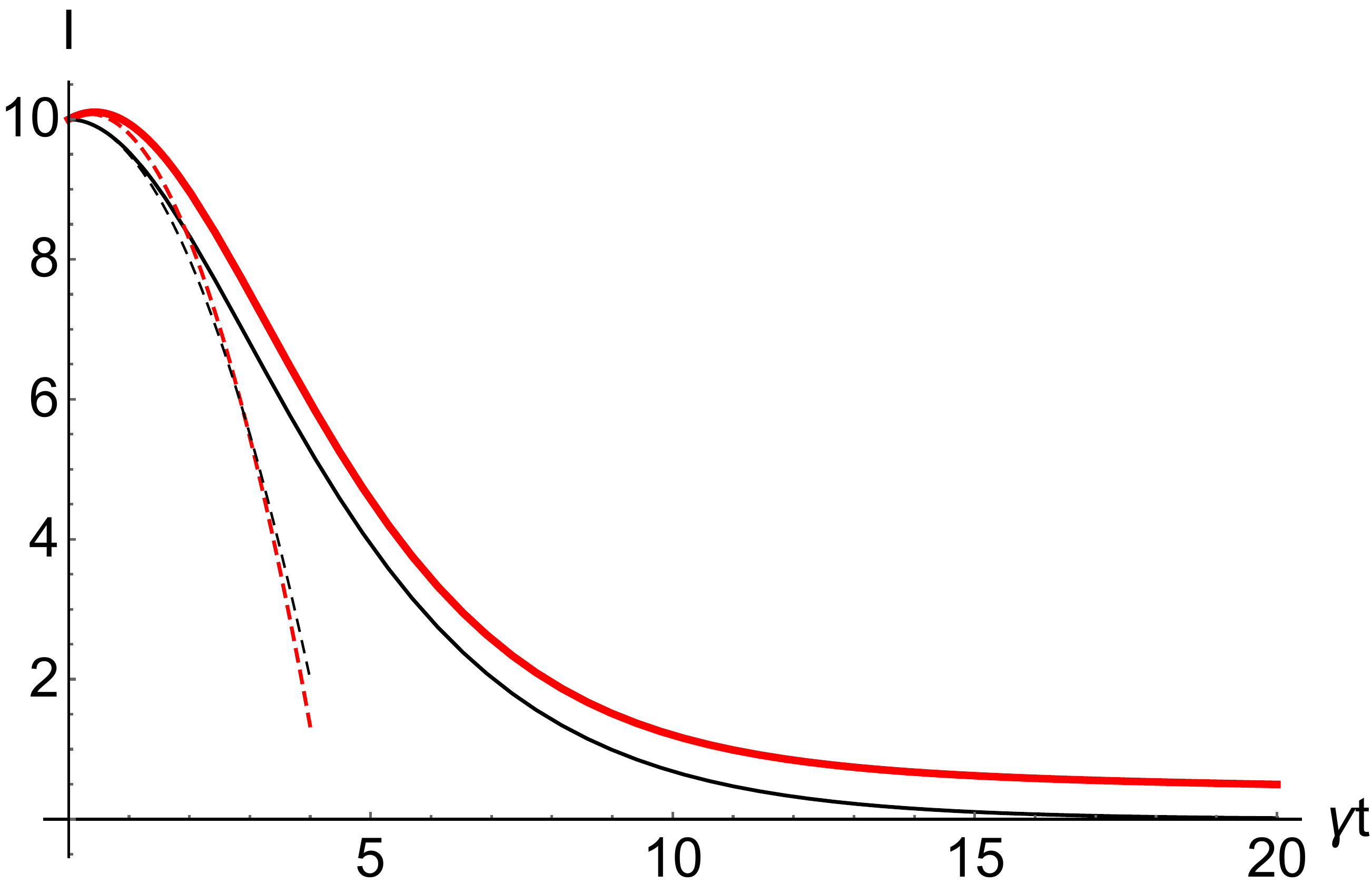}%
\includegraphics[keepaspectratio, scale=0.2]{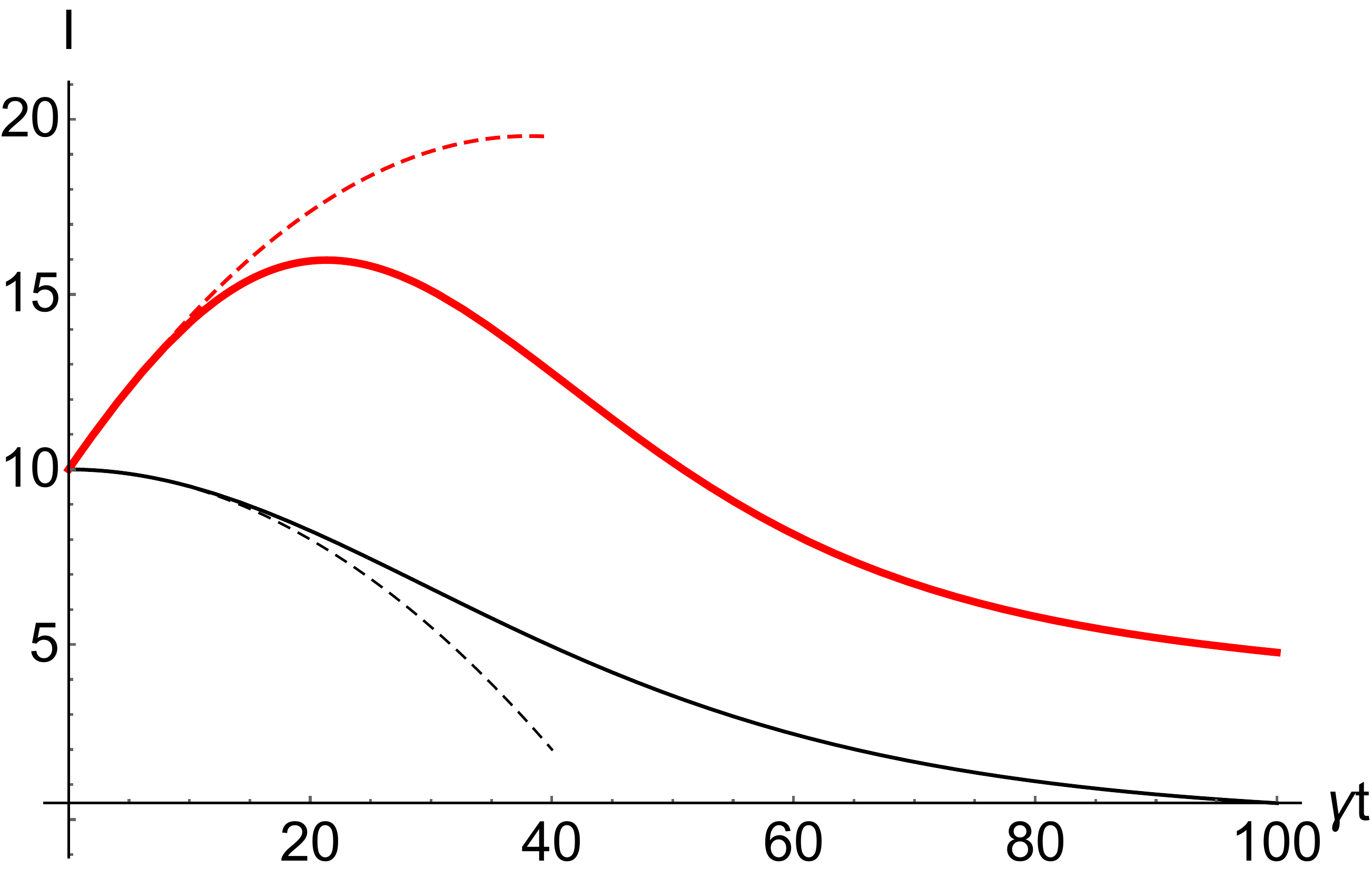}%
\includegraphics[keepaspectratio, scale=0.2]{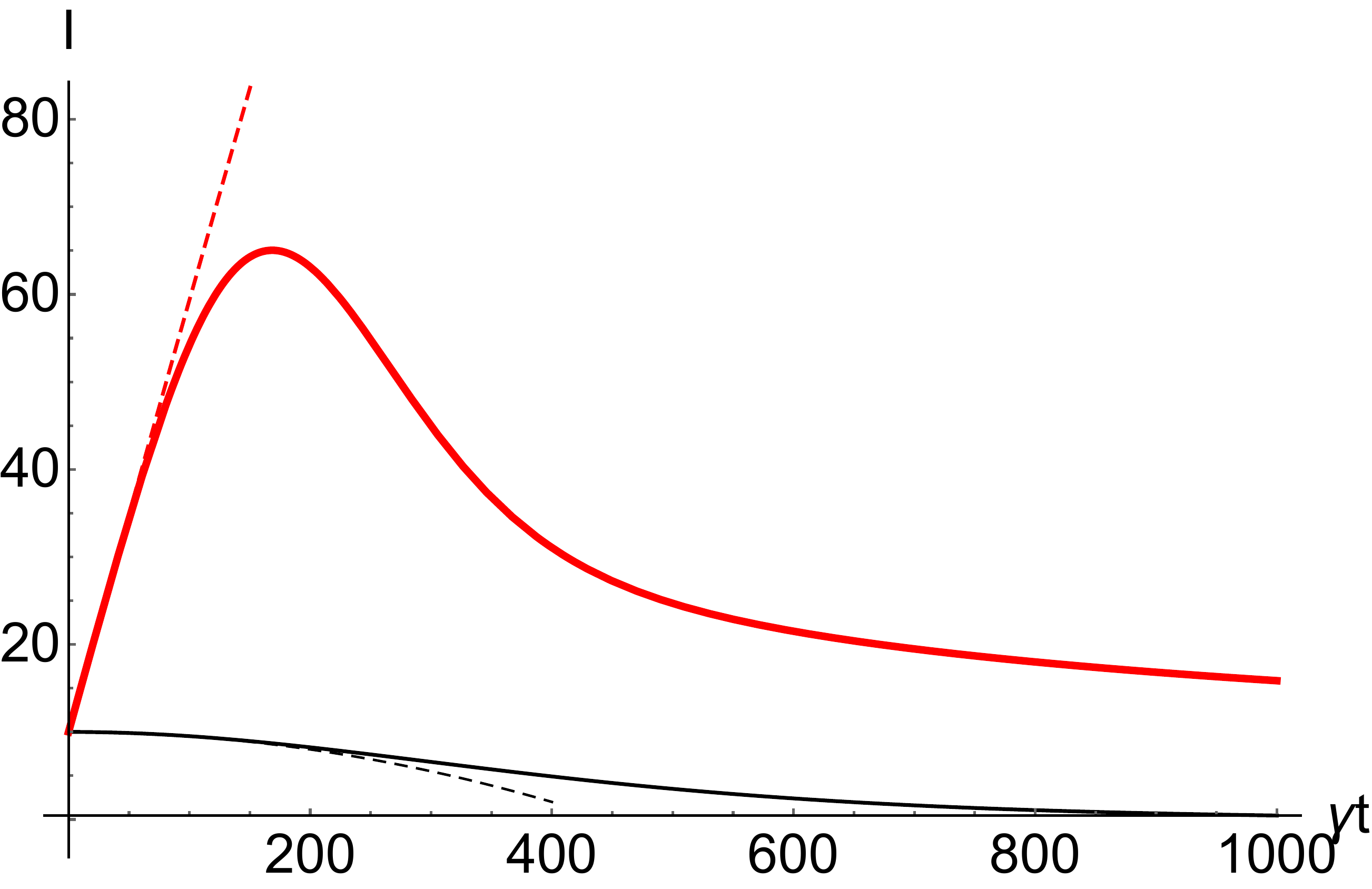}%
\\
\vspace*{2em}
\includegraphics[keepaspectratio, scale=0.2]{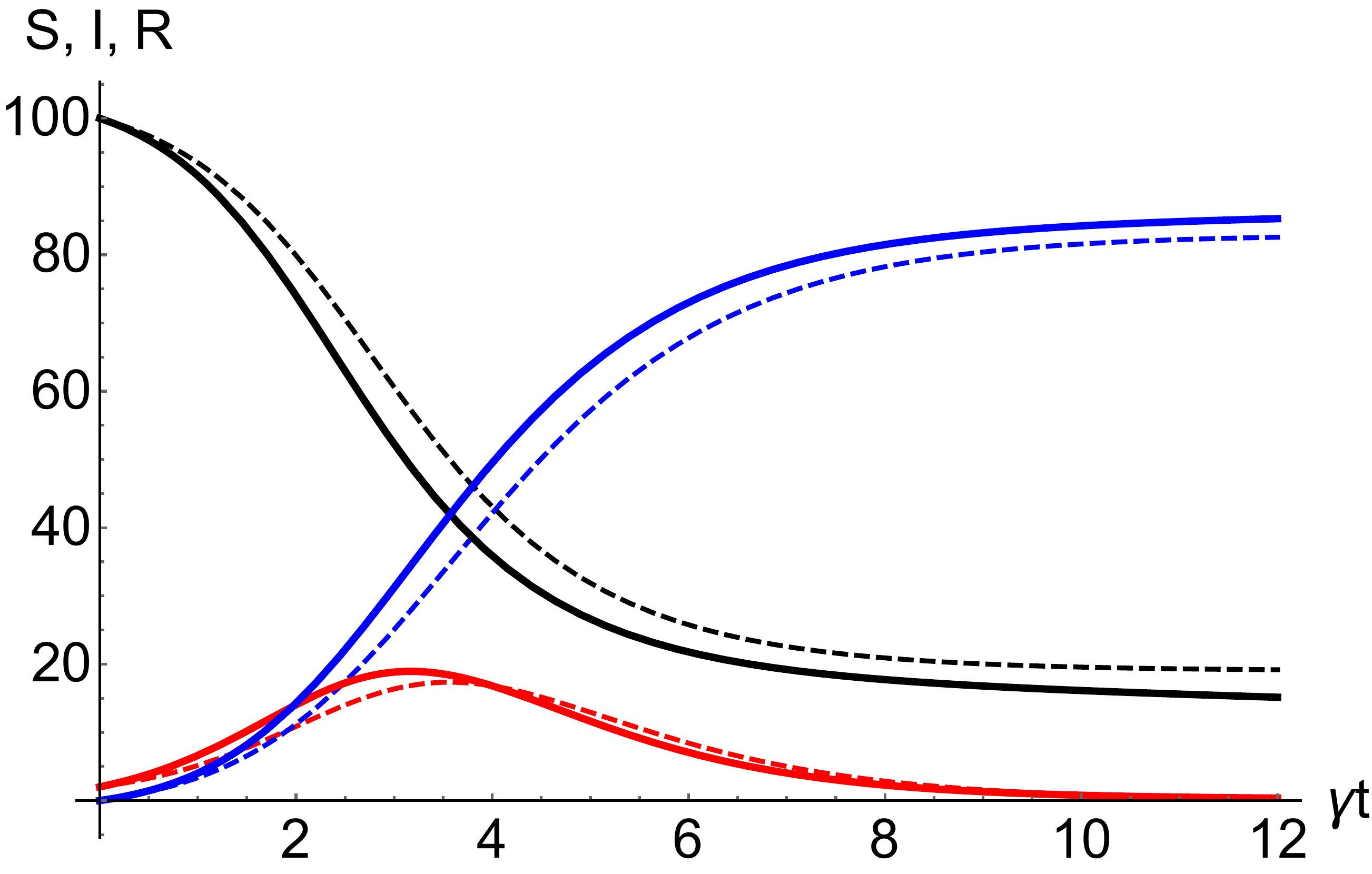}%
\includegraphics[keepaspectratio, scale=0.2]{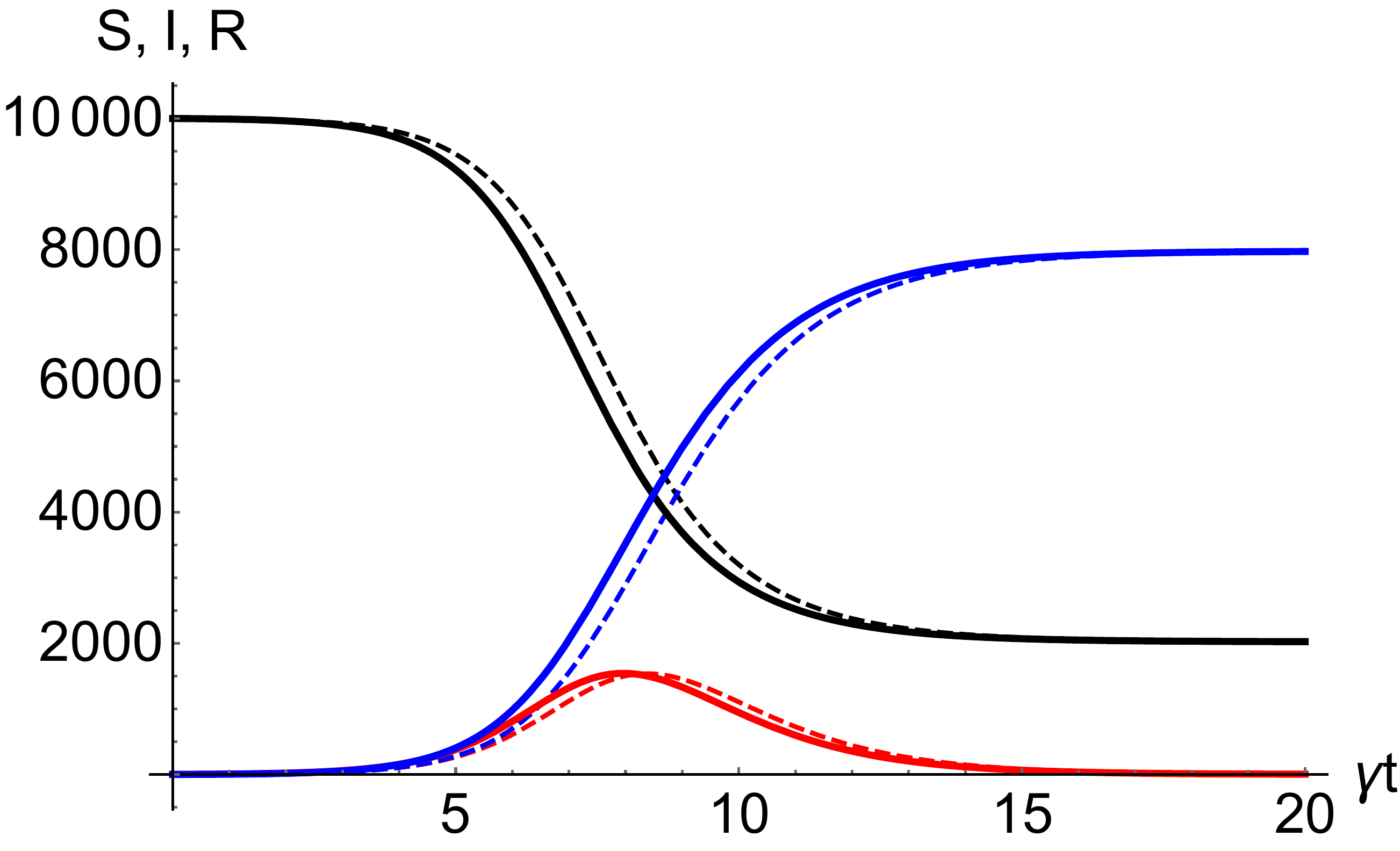}%
\includegraphics[keepaspectratio, scale=0.2]{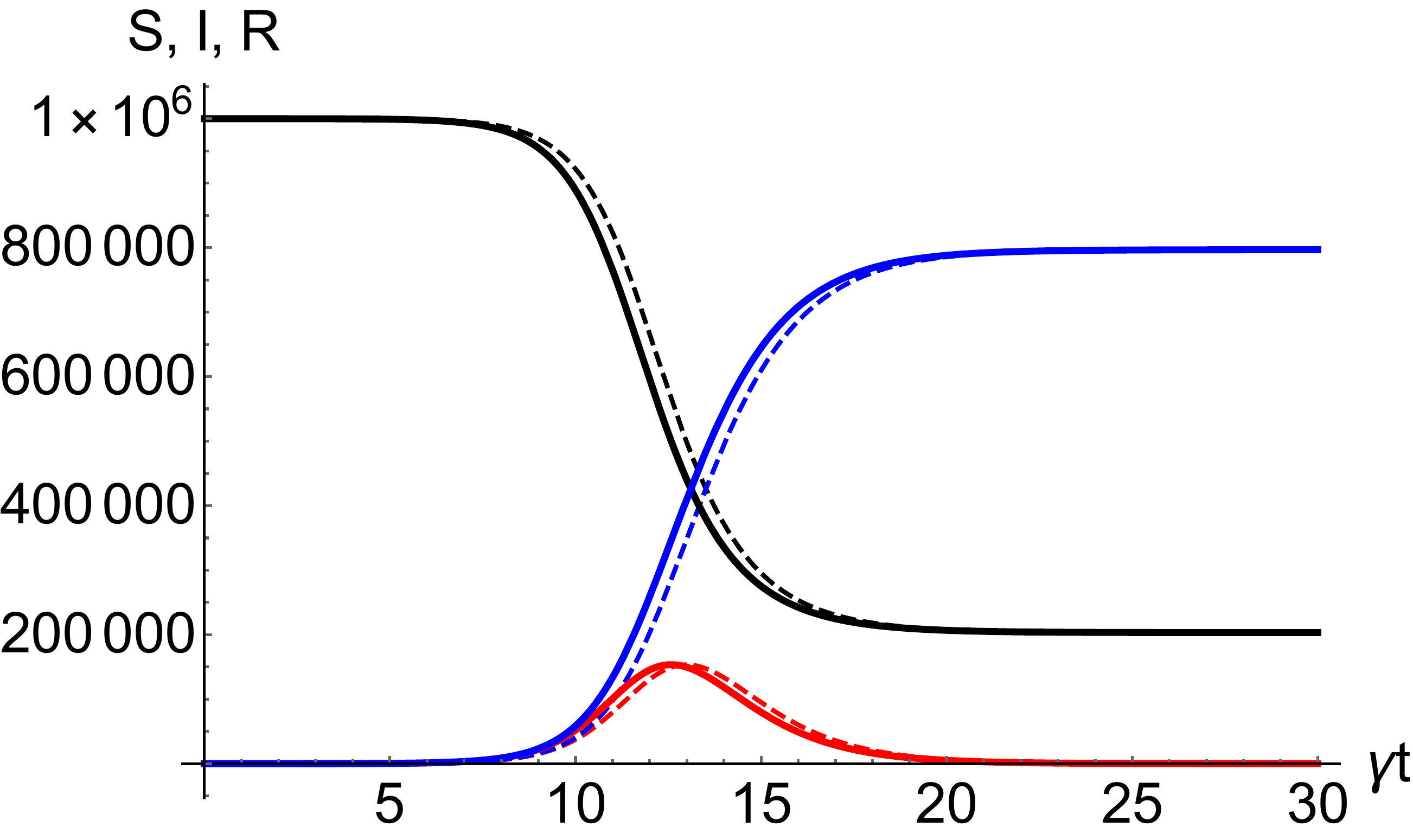}%
\caption{The numbers of susceptible, infectious, and removed individuals, $S(t)$, $I(t)$, and $R(t)$, are shown fo the classical and effective EL equations, Eqs.~\eqref{eq:SIR_model_differential_equation_boundary_S}-\eqref{eq:SIR_model_differential_equation_boundary_R} and Eqs.~\eqref{eq:SIR_EOM_fluctuation_limit_S}-\eqref{eq:SIR_EOM_fluctuation_limit_R}, respectively. The first, second, and third columns are for various reproduction numbers ${\cal R}=0.9$, $1$, and $2$, respectively. For ${\cal R}=0.9$ and $1$, only the results of infectious individuals are shown.
For each value of ${\cal R}$, the parameter sets of
$\bar{n}_{S0}=10^{2}$, $10^{4}$, and $10^{6}$ with common values $\bar{n}_{I0}=10$, $\bar{n}_{R0}=0$, and $\gamma=1$
are used from the left row to the right row.
In the top column, the horizontal red dashed lines indicates the asymptotic limit $I(\infty)=4.5$ in Eq.~\eqref{eq:SIR_NP_simple_one_loop_limit}.
In the top and middle columns, the black and red-solid lines are for the  classical and effective EL equations, and the black and red-dashed lines are for the perturbative results at classical and lone-loop levels, Eqs.~\eqref{eq:SIR_S_average_tree}-\eqref{eq:SIR_R_average_tree} and Eqs.~\eqref{eq:SIR_S_average}-\eqref{eq:SIR_R_average}, respectively.
In the bottom column, the solid and dashed lines are for the effective and classical EL equations, and the black, red, and blue lines are for the numbers of susceptible, infectious, and removed individuals.
\label{fig:SIR_fluctuation}}
\end{figure}

Finally, we show the numerical examples of the numbers of susceptible, infectious, and removed individuals, $S(t)$, $I(t)$, and $R(t)$,  for ${\cal R}=0.9$, $1$, and $2$ in Fig.~\ref{fig:SIR_fluctuation}.
We show the results from the classical and effective EL equations in Eqs.~\eqref{eq:SIR_model_differential_equation_boundary_S}-\eqref{eq:SIR_model_differential_equation_boundary_R} and Eqs.~\eqref{eq:SIR_EOM_fluctuation_limit_S}-\eqref{eq:SIR_EOM_fluctuation_limit_R}, respectively.
We also show the perturbative results at classical and one-loop levels in Eqs.~\eqref{eq:SIR_S_average_tree}-\eqref{eq:SIR_R_average_tree} and Eqs.~\eqref{eq:SIR_S_average}-\eqref{eq:SIR_R_average}, respectively.
For each value of ${\cal R}$, we use the parameter sets of
$\bar{n}_{S0}=10^{2}$, $10^{4}$, and $10^{6}$ for different population sizes with common values $\bar{n}_{I0}=10$, $\bar{n}_{R0}=0$, and $\gamma=1$, where the value of $\gamma$ gives the unit time of the recovery from infection.
As shown for ${\cal R}=2$ in the figure, we observe that the fluctuation effects are more important compared to the classical results when the population is small for the value of ${\cal R}$ far from the critical value.
Comparing the results of ${\cal R}=0.9$, $1$, and $2$, we find that the stochastic fluctuation becomes more enhanced for smaller ${\cal R}$ as a general tendency, as shown for ${\cal R}=0.9$ and $1$.
This result would be reasonable because the fluctuations near the critical point should be dominant rather than the mean values in general.
Besides, unexpectedly, we also find that the stochastic fluctuation becomes much more enhanced for larger number of initial susceptible individuals for the value of ${\cal R}$ near the critical value as shown for ${\cal R}=0.9$ and $1$.
This result seems against the naive expectation that the fluctuations are important in small-size systems.
Such behavior tells us the importance of the stochastic fluctuations not only in small-size systems but also in large-size systems  near the critical value.

We close our discussion by leaving some comments for Fig.~\ref{fig:SIR_fluctuation}.
For both ${\cal R}=0.9$ and $1$, it is shown that the perturbative results are consistent with the nonperturbative results at small $t$.
For ${\cal R}=0.9$, it is confirmed that
the asymptotic values of the infectious individuals approach the limit $I(\infty)=4.5$ in Eq.~\eqref{eq:SIR_NP_simple_one_loop_limit} and there remain several infectious people in the effective endemic state, when the conditions $I(t) \ll S(t)$ and ${\cal R}<1$ are satisfied in the stochastic SIR model.
For ${\cal R}=1$, we may wonder why the numbers of infectious individuals increase at small $t$. This is because the critical value of the basic reproduction number becomes smaller than one by the stochastic fluctuations in the perturbative case as shown in Sec.~\ref{sec:perturbation_tree}, see Eq.~\eqref{eq:SIR_I_average_approx}.

\section{Conclusion and perspectives} \label{sec:conclusion}

We have discussed the critical value of the basic reproduction number in the stochastic SIR model for infectious diseases. We have formulated the master equation for the stochastic process
 among susceptible, infectious, and removed individuals and have rewritten it in terms of the Hamiltonian formalism, and finally have transformed it into the path-integral formalism.
This is analogous to the formulation in quantum field theory for microscopic physics.
Based on the generating functional, we have performed the perturbative and nonperturbative analyses to evaluate the critical value of the basic reproduction number, ${\cal R}$.
In the perturbation theory, we have calculated the mean values and the variances for the numbers of susceptible, infectious, and removed individuals as a series of time near the initial time.
We have found that the critical value ${\cal R}_{\c}=1/3$--$2/3$ should be adopted in order to suppress the stochastic spread of infections certainly within the probabilistic uncertainty.
In the nonperturbative approach, we have derived the effective potential by integrating out the stochastic fluctuations at one-loop level,
 and have obtained the effective Euler-Lagrange equations for the time-evolution of the numbers of susceptible, infectious, and removed individuals.
The effective Euler-Lagrange equations include the new terms generated by the stochastic fluctuations which are absent in the classical SIR model.
Assuming that the number of susceptible individuals is much larger than that of infectious individuals,
we have found that the critical value ${\cal R}_{\c}=2/3$ should be adopted for the complete convergence of infections.
As a conclusion, the critical value of the basic reproduction number should be less than one, against the known critical value ${\cal R}_{\c}=1$ in the classical SIR model, when the stochastic fluctuations are taken into account.

In the present study, we have introduced the bosonic commutation relations in the Hamiltonian formalism based on the assumption that there can be infinitely many individuals.
In reality, however, there should be a maximum number of individuals.
Such restriction can be described in terms of the fermion operators as well as the parastatistics operators as an extension from the boson operators.
We have assumed that the coefficients in the stochastic SIR model are constant number which should be generalized to the time-dependent function for application to real-world data.
The age structures should also be discussed together with the extension to the SEIR and SEIRS models including the treatment and hospitalization effects.
In order to follow up the activity of each individual, the network epidemiology is also an interesting topics which can be applied by extending present analysis of the stochastic SIR model.
These subjects are left for future research.

\appendix

\section{Generating functional} \label{sec:generating_functional}

We can obtain the Feynman rules systematically from the generating functional~\eqref{eq:SIR_generating_functional_source_term_def}.
For this purpose, we transform Eq.~\eqref{eq:SIR_generating_functional_source_term_def} to 
\begin{align}
   Z[\bar{j},j]
=
   {\cal N}
   \exp
   \Bigl(
         {\partial}_{\phi} \Delta {\partial}_{\bar\phi}
      + \bar{n}_{0} {\partial}_{\phi}
   \Bigr)
   \exp
   \Bigl(
       - S^{\intn}[\bar{\phi},\phi]
      + \bar{\phi}j + \bar{j}\phi
   \Bigr)
   \biggr|_{\phi=\bar{\phi}=0},
\label{eq:SIR_generating_functional_source_term_2}
\end{align}
where we use the following notations
\begin{align}
   {\partial}_{\phi} \Delta {\partial}_{\bar\phi}
&\equiv
   {\partial}_{\rS} \Delta {\partial}_{\bar{S}}
+ {\partial}_{\rI} \Delta {\partial}_{\bar{I}}
+ {\partial}_{\rR} \Delta {\partial}_{\bar{R}},
\\ 
   \bar{n}_{0} {\partial}_{\phi}
&\equiv
   \bar{n}_{S0} {\partial}_{\rS}
+ \bar{n}_{I0} {\partial}_{\rI}
+ \bar{n}_{R0} {\partial}_{\rR},
\\ 
   \bar{\phi}j
&\equiv
   \bar{S}\sigma
+ \bar{I}\iota
+ \bar{R}\rho,
\\ 
   \bar{j}\phi
&\equiv
   \bar{\sigma}S
+ \bar{\iota}I
+ \bar{\rho}R,
\end{align}
and
\begin{align}
   {\partial}_{\phi} \equiv \frac{\delta}{\delta\phi},
\quad 
   {\partial}_{\bar{\phi}} \equiv \frac{\delta}{\delta\bar{\phi}},
\quad 
   \Delta \equiv {\dr}_{\tau}^{-1},
\quad 
   {\dr}_{\tau}\equiv\frac{{\dr}}{{\dr}\tau},
\end{align}
for $\phi=\bigl\{S,I,R\bigr\}$ and $\bar{\phi}=\bigl\{\bar{S},\bar{I},\bar{R}\bigr\}$.
For brevity, we have introduced the abbreviations
\begin{align}
   {\partial}_{\phi} \Delta {\partial}_{\bar{\phi}}
=
   \int_{0}^{t} \dr \tau \, \frac{\delta}{\delta\phi} {\dr}_{\tau}^{-1} \frac{\delta}{\delta\bar{\phi}},
\quad 
   \bar{n}_{0} {\partial}_{\phi},
=
   \bar{n}_{0} \int_{0}^{t} \dr \tau \frac{\delta}{\delta\phi},
\end{align}
and
\begin{align}
   \bar{\phi}j+\bar{j}\phi
=
   \int_{0}^{t} \dr \tau \, (\bar{\phi}j+\bar{j}\phi).
\end{align}
The equation~\eqref{eq:SIR_generating_functional_source_term_2} can be obtained in the following way.
First, we transform the generating functional~\eqref{eq:SIR_generating_functional_source_term_def} to
\begin{align}
   Z[\bar{j},j]
= {\cal N}
   G\biggl[\frac{\delta}{\delta\bar{j}},\frac{\delta}{\delta{j}}\biggr]
   F[\bar{j},j],
\label{eq:generating_functional_1}
\end{align}
with the functionals defined by
\begin{align}
   G\biggl[\frac{\delta}{\delta{j}},\frac{\delta}{\delta\bar{j}}\biggr]
&\equiv
   \exp
   \biggl(
        - S_{\intn}\biggl[\frac{\delta}{\delta j},\frac{\delta}{\delta \bar{j}}\biggr]
   \biggr),
\\ 
   F[\bar{j},j]
&\equiv
   \exp
   \biggl(
         \int_{0}^{t} \dr \tau \, \bar{j} {\dr}_{\tau}^{-1} j
      + \bar{n}_{0} \int_{0}^{t} \dr \tau \, \bar{j}
   \biggr).
\end{align}
Using that the functionals $F$ and $G$ satisfy the following relation
\begin{align}
   G\biggl[\frac{\delta}{\delta{j}},\frac{\delta}{\delta\bar{j}}\biggr]
   F[\bar{j},j]
&=
   F\biggl[\frac{\delta}{\delta\phi},\frac{\delta}{\delta\bar{\phi}}\biggr]
   G[\bar{\phi},\phi]
   \exp\biggl(\int_{0}^{t} \dr \tau \, (\bar{\phi}j+\bar{j}\phi) \biggr)
   \biggr|_{\phi=\bar{\phi}=0},
\end{align}
we further transform Eq.~\eqref{eq:generating_functional_1} to
\begin{align}
   Z[\bar{j},j]
&=
   {\cal N}
   F\biggl[\frac{\delta}{\delta\phi},\frac{\delta}{\delta\bar{\phi}}\biggr]
   G[\bar{\phi},\phi]
   \exp\biggl(\int_{0}^{t} \dr \tau \, (\bar{\phi}j+\bar{j}\phi) \biggr)
   \biggr|_{\phi=\bar{\phi}=0}.
\end{align}
Finally, we obtain the expression in Eq.~\eqref{eq:SIR_generating_functional_source_term_2}.

\section{Expectation values and variances} \label{sec:expectation_values_variances}

The mean values and variances for irreducible diagrams in Sec.~\ref{sec:generating_functional} are derived by using
\begin{align}
   \frac{\delta}{\delta \bar{j}_{\tau}} W[\bar{j},j] \biggr|_{\bar{j}=j=0}
&=
   \frac{1}{Z[\bar{j},j]}
   \frac{\delta}{\delta \bar{j}_{\tau}} Z[\bar{j},j] \biggr|_{\bar{j}=j=0},
\end{align}
and
\begin{align}
   \frac{\delta^{2}}{\delta \bar{j}_{\tau} \delta \bar{j}_{\sigma}}
   W[\bar{j},j] \biggr|_{\bar{j}=j=0}
&=
   \frac{1}{Z[\bar{j},j]}
   \frac{\delta^{2}}{\delta \bar{j}_{\tau} \delta \bar{j}_{\sigma}} Z[\bar{j},j]
   \Biggr|_{\bar{j}=j=0}
 - \frac{1}{Z[\bar{j},j]^{2}}
   \biggl( \frac{\delta}{\delta \bar{j}_{\tau}} Z[\bar{j},j] \biggr)
   \biggl( \frac{\delta}{\delta \bar{j}_{\sigma}} Z[\bar{j},j] \biggr)
   \Biggr|_{\bar{j}=j=0},
\end{align}
from the generating functional $Z[\bar{j},j]$ with source functions $j$ and $\bar{j}$.
The expressions of the variances~\eqref{eq:number_fluctuation_value_SIR} are obtained through
$\langle1|(\hat{a}^{\dag}\hat{a})^{2}|\phi_{t}\rangle = \langle1| (\hat{a}+1) \hat{a} |\phi_{t}\rangle$ for the operator $(\hat{a}^{\dag}\hat{a})^{2}$ compound by the annihilation and creation operators,
indicating that $(\hat{a}^{\dag}\hat{a})^{2}$ corresponds to $\bigl(\phi(t)+1\bigr)\phi(t)$.

\section{Calculation of effective potential} \label{sec:effective_potential_calc}

We calculate the trace term in Eq.~\eqref{eq:SIR_effective_potential_1loop} in the following way:
\begin{align}
   \Tr \ln \Delta^{-1}[\bar{\phi},\phi]
&=
   \tr
   \int_{0}^{t} \dr \tau \, \ln \Delta^{-1}[\bar{\phi},\phi](\tau,\tau)
\nonumber \\ 
&=
   \tr
   \int_{0}^{t} \dr \tau \,
   \langle \tau |
   \Biggl(
         \ln
         \bigl( {\dr}_{\tau} - {\cal M}_{\c}(\tau) \bigr)
      + \ln
         \biggl(
               1
            + \frac{\delta(\tau)}{{\dr}_{\tau} - {\cal M}_{\c}(\tau)}
         \biggr)
   \Biggr)
   | \tau \rangle
\nonumber \\ 
&=
   \tr
   \int_{0}^{t} \dr \tau \,
   \langle \tau |
   \ln \bigl( {\dr}_{\tau} - {\cal M}_{\c}(\tau) \bigr)
   | \tau \rangle
+ \tr
   \int_{0}^{t} \dr \tau \,
   \langle \tau |
   \ln
   \biggl(
         1
      + \frac{\delta(\tau)}{{\dr}_{\tau} - {\cal M}_{\c}(\tau)}
   \biggr)
   | \tau \rangle
\nonumber \\ 
&=
   \frac{1}{2}
   \int_{0}^{t} \dr \tau \,
   \sum_{\alpha=\pm,0}
   \int_{0}^{m_{\c\alpha}(\tau)} \dr x \,
   \coth \biggl( \frac{t}{2}x \biggr)
+ \sum_{\alpha=\pm,0}
   \ln
   \Biggl(
         1
       - \frac{1}{2} \coth \biggl( \frac{t}{2}m_{\c\alpha}(0) \biggr)
   \Biggr),
\label{eq:SIR_effective_trace_calculation_result}
\end{align}
where we sum over the Matsubara frequencies $\omega_{n}=2\pi n t$ ($n \in {\mathbb Z}$).
Here we have introduced the Fourier transformation whose basis is spanned by $|\omega_{n}\rangle$, and have used the product $\langle \omega_{n} | \tau \rangle=e^{i\omega_{n}\tau}$ represented by the plane-wave,
 the completeness relation
\begin{align}
   \frac{1}{t} \sum_{n} |\omega_{n}\rangle \langle\omega_{n}|=1,
\end{align}
and the summation over the Matsubara frequencies
\begin{align}
   \sum_{n}
   \frac{1}{i\omega_{n}+x}
=
   \frac{t}{2}
   \coth \biggl( \frac{t}{2}x \biggr),
\end{align}
for an arbitral real number $x$.
In the above calculation, we have used the eigenvalues in the matrix ${\cal M}_{\c}$ in Eq.~\eqref{eq:SIR_Mc_def}, 
\begin{align}
   m_{\c \pm}(\tau)
&\equiv
   \frac{1}{2} 
   \Biggl(
         \beta \bigl(2\bar{I}_{\c}-\bar{S}_{c}+1\bigr) S_{\c}
       - \beta \bigl(\bar{I}_{\c}+1\bigr) I_{\c}
       - \gamma
         \nonumber \\ & \hspace{2em} 
  \pm \sqrt{
               \Bigl(
                     \beta \bigl(2\bar{I}_{\c}-\bar{S}_{c}+1\bigr) S_{\c}
                  + \beta \bigl(\bar{I}_{\c}+1\bigr) I_{\c}
                   - \gamma
               \Bigr)^{2}
            - 4
              \beta^{2}
              \bigl(\bar{I}_{\c}+1\bigr)
              \bigl(2\bar{I}_{\c}-\bar{S}_{\c}+1\bigr)
              S_{\c}
              I_{\c}
         }
   \Biggr),
\\ 
   m_{\c 0}(\tau) &\equiv 0.
\end{align}
Inserting Eq.~\eqref{eq:SIR_effective_trace_calculation_result} into Eq.~\eqref{eq:SIR_effective_potential_1loop},
 we finally obtain the approximate form of the effective potential for arbitrary $t$:
\begin{align}
   \Gamma[\bar{\phi},\phi]
&\approx
 - \int_{0}^{t} \dr \tau
   \Biggl(
               \bar{S} \frac{{\dr}}{{\dr} \tau} S
            + \bar{I} \frac{{\dr}}{{\dr} \tau} I
            + \bar{R} \frac{{\dr}}{{\dr} \tau} R
            + \beta \bigl( - \bar{I}^{2} + \bar{S} \bar{I} - \bar{I} + \bar{S} \bigr) S I
            + \gamma \bigl( - \bar{R} + \bar{I} \bigr)I
             - \frac{1}{4}
               \sum_{\alpha=\pm}
               \int_{0}^{m_{\c\alpha}(\tau)} \dr x \,
               \coth \biggl( \frac{t}{2}x \biggr)
               \nonumber \\ & \hspace{5em} 
            + \Biggl(
                     \bar{S} \bigl(S-\bar{n}_{S0}\bigr)
                  + \bar{I} \bigl(I-\bar{n}_{I0}\bigr)
                  + \bar{R} \bigl(R-\bar{n}_{R0}\bigr)
                  + \frac{1}{2}
                     \sum_{\alpha=\pm}
                     \ln
                     \Biggl(
                           1
                         - \frac{1}{2} \coth \biggl( \frac{t}{2}m_{\c\alpha}(\tau) \biggr)
                     \Biggr)
               \Biggr)
               \delta(\tau)
   \Biggr).
\label{eq:SIR_effective_potential_1}
\end{align}

\bibliography{reference}

\end{document}